\documentclass[12pt,english, journal,draftclsnofoot,onecolumn]{IEEEtran}
\usepackage[T1]{fontenc}
\usepackage[latin9]{inputenc}
\usepackage{color}
\usepackage{float}
\usepackage{textcomp}
\usepackage{mathrsfs}
\usepackage{amsthm}
\usepackage{amsmath}
\usepackage{amssymb}
\usepackage{stackrel}
\usepackage{graphicx}
\usepackage{setspace}
\usepackage{esint}
\makeatletter

\floatstyle{ruled}
\newfloat{algorithm}{tbp}{loa}
\providecommand{\algorithmname}{Algorithm}
\floatname{algorithm}{\protect\algorithmname}

\theoremstyle{plain}
\newtheorem{thm}{\protect\theoremname}

\usepackage{colortbl}
\usepackage{cite}
\usepackage{nomencl}
\usepackage{flushend}
\usepackage{stfloats}
\usepackage[caption=false]{subfig}

\@ifundefined{showcaptionsetup}{}{%
 \PassOptionsToPackage{caption=false}{subfig}}
\usepackage{subfig}
\makeatother

\usepackage{babel}
\providecommand{\theoremname}{Theorem}

\begin{document}

\title{STAR-RIS Assisted Full-Duplex Communication Networks}

\author{Abdelhamid Salem,\textit{\normalsize{} Member, IEEE}{\normalsize{},}
and Kai-Kit Wong, \textit{\normalsize{}Fellow, IEEE}, Chan-Byoung
Chae,\textit{\normalsize{} Fellow, IEEE} and Yangyang Zhang \\
\thanks{Abdelhamid Salem is with the department of Electronic and Electrical
Engineering, University College London, London, UK, (emails: a.salem@ucl.ac.uk).

Kai-Kit Wong is with the department of Electronic and Electrical Engineering,
University College London, London, UK, (email: kai-kit.wong@ucl.ac.uk).
Kai-Kit Wong is also affiliated with Yonsei University, Seoul, Korea. 

Chan-Byoung Chae is with the School of Integrated Technology, Yonsei
University, Seoul, 03722 Korea (e-mail: cbchae@yonsei.ac.kr). 

Yangyang Zhang is with Kuang-Chi Science Ltd., Hong Kong, SAR, China.%
} }
\maketitle
\begin{abstract}
Different from conventional reconfigurable intelligent surfaces (RIS),
a recent innovation called simultaneous transmitting and reflecting
reconfigurable intelligent surface (STAR-RIS) has emerged, aimed at
achieving complete 360-degree coverage in communication networks.
Additionally, full-duplex (FD) technology is recognized as a potent
approach for enhancing spectral efficiency by enabling simultaneous
transmission and reception within the same time and frequency resources.
In this study, we investigate the performance of a STAR-RIS-assisted
FD communication system. The STAR-RIS is strategically placed at the
cell-edge to facilitate communication for users located in this challenging
region, while cell-center users can communicate directly with the
FD base station (BS). We employ a non-orthogonal multiple access (NOMA)
pairing scheme and account for system impairments, such as self-interference
at the BS and imperfect successive interference cancellation (SIC).
We derive closed-form expressions for the ergodic rates in both the
up-link and down-link communications and extend our analysis to bidirectional
communication between cell-center and cell-edge users. Furthermore,
we formulate an optimization problem aimed at maximizing the ergodic
sum-rate. This optimization involves adjusting the amplitudes and
phase-shifts of the STAR-RIS elements and allocating total transmit
power efficiently. To gain deeper insights into the achievable rates
of STAR-RIS-aided FD systems, we explore the impact of various system
parameters through numerical results.\end{abstract}


\section{Introduction}

Reconfigurable intelligent surface (RIS) has been recognized as a
key technique for the upcoming sixth-generation (6G) wireless communication
networks \cite{ref1,Reef1,https://doi.org/10.48550/arxiv.2301.00276,Dai20}.
The conventional RIS is equipped with controllable reflecting elements,
that can adjust the phase shifts of the incident signals to improve
the received signals' quality \cite{ref1,Reef1,https://doi.org/10.48550/arxiv.2301.00276}.
Therefore, RIS can make the propagation environments smart and controllable.
The performance of RIS has been extensively studied in the literature
for different applications \cite{Ref22,Ref33,Ref44,Ref55}. However,
the conventional RIS can provide only half-space coverage, and thus
the transmitters and receivers should be located on the same side
of the aided RIS. To tackle this issue and extend the RIS coverage
area, simultaneous transmitting and reflecting RIS (STAR-RIS) has
been proposed recently \cite{refone,mag1,Ref2,Ref3,Ref4,Ref5,Ref6,Ref7,Ref8,Ref9,ref9a,ref9b}.
STAR-RIS can transmit and reflect the signals into both sides of the
surface, and thus it can provide full-space coverage. The concept
of STAR-RIS was introduced and discussed in \cite{refone,mag1} where
three operating protocols for the STAR-RIS have been proposed, namely,
energy splitting (ES), mode switching (MS), and time switching (TS).
In ES, the energy of the incident wave on each STAR-RIS element is
split into energy of the transmitted signals and energy of the reflected
signals, and in the MS protocol, the STAR-RIS elements are divided
into two groups: one group operates in transmission mode, while the
other group operates in reflecting mode. Also, in the TS protocol,
the STAR-RIS elements periodically switch between transmission mode
and reflection mode in different time slots. The authors in \cite{Ref2}
analyzed the ergodic rates of the STAR-RIS-assisted non-orthogonal
multiple access (NOMA) systems, where a STAR-RIS was implemented to
assist the cell-edge users. In \cite{Ref3}, the performance of STAR-RIS-aided
NOMA systems was studied to support low-latency communications. In
\cite{Ref4}, an analytical expression of the coverage probability
for a STAR-RIS aided massive multiple input multiple output (mMIMO)
communication systems was derived. Further work in \cite{Ref5} provided
an analytical framework of the coverage probability and ergodic rate
of STAR-RIS assisted NOMA multi-cell networks. A power minimization
problem for STAR-RIS-aided up-link NOMA systems was considered in
\cite{Ref6}. The authors in \cite{Ref7} considered the energy efficiency
of a STAR-RIS-empowered MIMO-NOMA systems. In \cite{Ref8}, an approximate
analytical expression of the ergodic rate of STAR-RIS-aided NOMA downlink
communication system was derived. A new joint optimization problem
to maximize the achievable sum rate of STAR-RIS NOMA system has been
formulated and solved in \cite{Ref9}. In \cite{ref9a}, a STAR-RIS-assisted
two-user communication systems has been considered for both orthogonal
multiple access (OMA) and NOMA schemes. The performance of STAR-RIS
aided NOMA systems over Rician fading channels has been analyzed in
\cite{ref9b}.

Moreover, full-duplex (FD) technique allows the communication nodes
to transmit and receive data simultaneously in the same frequency
and time resources \cite{Chae1,Chae2,Chae3}. Thus, FD can enhance
the achievable rates and provide a more flexible use of the spectrum.
Interestingly, the performance of RIS-aided FD communication systems
has been considered recently in several works \cite{Ref12,Ref13,ref14,ref15,Ref17,b1,b2}.
For instance, in \cite{Ref12}, a passive beamforming design for RIS-assisted
FD communication was investigated, where a FD access point communicates
with an uplink (UL) user and a downlink (DL) user simultaneously with
the help of RIS. In \cite{Ref13}, the resource allocation design
for RIS-assisted FD communication systems was considered. In \cite{ref14},
the authors proposed deploying an RIS in a FD two-way communication
systems to provide signal coverage for the users in the dead areas.
A joint beamforming design for a RIS-assisted FD communication systems
was considered in \cite{ref15}, where the total transmit power was
minimized by optimizing the active beamforming at the transmitter
and passive beamforming at the RIS. In \cite{Ref17} a RIS-aided FD
communication system has been analyzed, where a FD-BS communicates
with FD-users through a dedicated RIS.

Accordingly, this work considers a STAR-RIS aided FD communication
system, where the FD-BS serves the cell-center users and cell-edge
users using NOMA pairing scheme. The BS communicates with the cell-edge
users via passive STAR-RIS, while the cell-center users can communicate
directly with the BS. We concentrate on the ergodic rate analysis,
the STAR-RIS design and the power allocation, by employing statistical
channel state information (CSI). We first derive closed-form expressions
for the ergodic rates of the DL and UL users in the system. Then,
we apply our analysis to the bidirectional communication between the
cell-center and cell-edge users. In addition, the sum-rate of both
scenarios are maximized by optimizing the STAR-RIS reflection coefficients
and the power allocation between the DL and UL users. For clarity,
the main contributions are listed as follows: 

1) We investigate the performance of STAR-RIS aided FD-BS communication
systems, where the STAR-RIS is implemented to assist the cell-edge
users.

2) New closed-form analytical expressions for the ergodic rates of
the DL and UL users are derived when the cell-center users are paired
with the cell-edge users under Rician fading channels. This channel
model is more general but also very challenging to analyze. Also,
the impact of imperfect successive interference cancellation (SIC)
and all interferences including the self interference at the FD-BS
are considered in the analysis. The derived ergodic rate expressions
are explicit and provide several important practical design insights. 

3) Besides many applications, our analysis is then applied to the
bidirectional communication scenario, where the UL cell-center user
and UL cell-edge user communicate with the DL cell-edge user and DL
cell-center user, respectively. Simple closed form expressions for
the ergodic rates of the DL users are derived.

4) We formulate and solve a sum-rate maximization problem by jointly
optimizing the phase shifts and amplitudes at the STAR-RIS and the
power allocation coefficients. To efficiently solve this challenging
problem, we divide the main problem into two sub-problems, i.e., phase
shift and amplitude optimization, and power allocation optimization,
then we solve them alternatively. In addition, based on the derived
rate expressions, we introduce sub-optimal designs of the STAR-RIS
and the power allocation schemes.

5) Monte-Carlo simulations are performed to validate the analytical
expressions. Then, the impact of several parameters on the system
performance are investigated. 

The results in this work show that increasing the transmit signal
to noise ratio (SNR) always enhances the achievable users' rates,
and using a large number of STAR-RIS units improves the performance
of the cell-edge users. In addition, high power should be allocated
to the DL users to overcome the interference caused by the UL users.
The imperfect SIC degrades the achievable rates of the DL cell-center
user and the UL cell-edge user, while\textcolor{red}{{} }a high variance
of the residual self interference at the BS leads to degrade the performance
of the UL users significantly. 

Next, in Section \ref{sec:System-Model} the system model is presented.
In Section \ref{sec:NOMA-Pairing}, we derive the ergodic rates of
the DL and UL users in the NOMA pairing scheme. The ergodic rates
of the DL users in the bidirectional communication scenario are provided
in Section \ref{sec:Bidirectional-Communication}. Section \ref{sec:System-Design}
discusses the optimal system design. Section \ref{sec:Numerical-Results}
depicts our numerical results. Our main conclusions are summarized
in Section \ref{sec:CONCLUSIONs}.

\section{System Model\label{sec:System-Model}}

\begin{figure}
\noindent \begin{centering}
\includegraphics[bb=77bp 10bp 780bp 500bp,clip,scale=0.33]{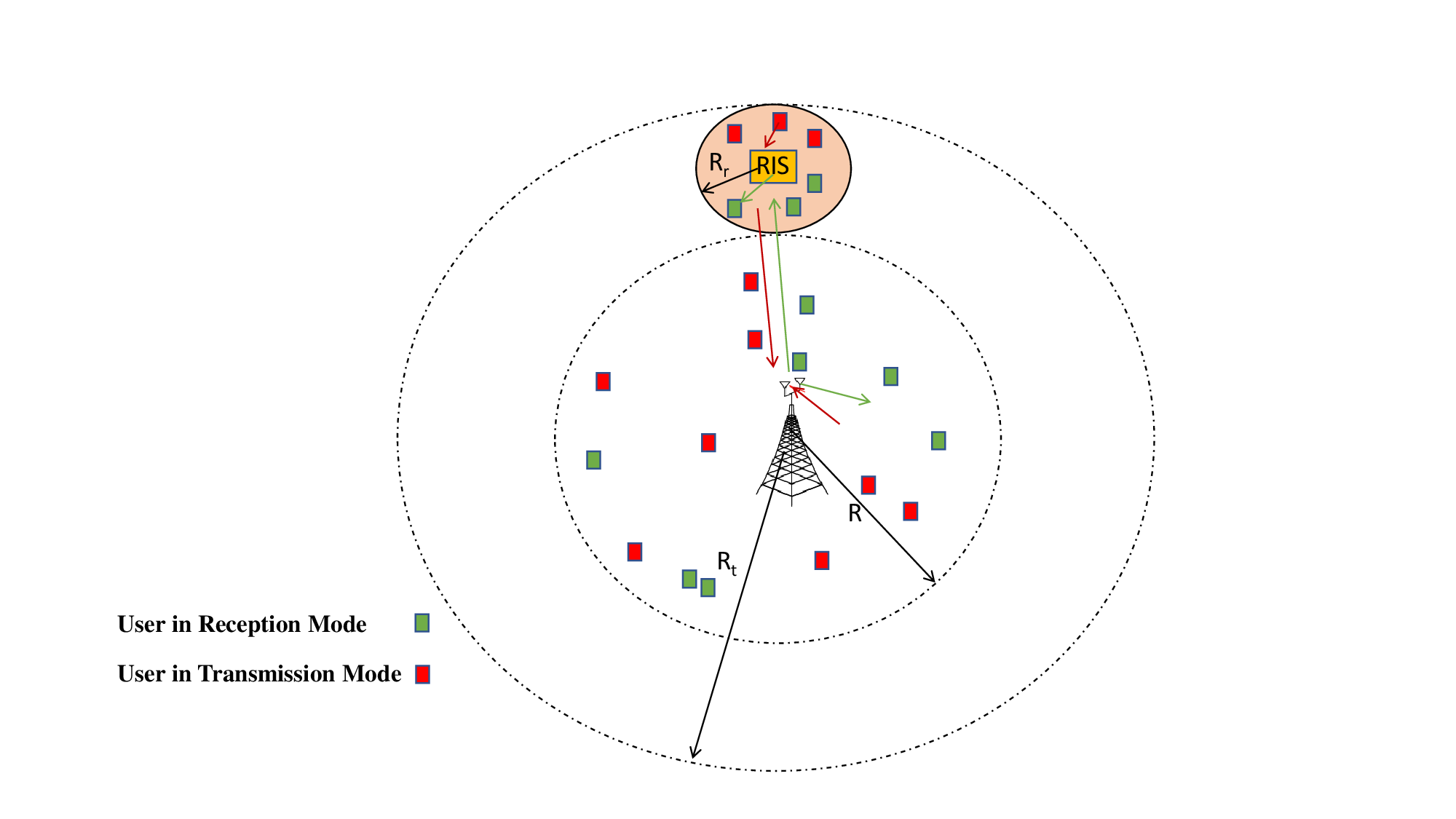}
\par\end{centering}

\protect\caption{\label{fig:System-Model}A STAR-RIS assisted FD communication system.}
\end{figure}

We consider a STAR-RIS-aided multiuser communication system operating
in FD mode, as shown in Fig \ref{fig:System-Model}. The FD-BS is
deployed at the cell-center with a coverage radius $R_{t}$, while
a STAR-RIS with $N$ reconfigurable elements is deployed at the cell-edge
with a coverage radius $R_{r}$. Based on the BS and the STAR-RIS
deployments, the total coverage area $R_{t}$ is divided into two
areas, i.e., the cell-center region with radius $R$, and the cell-edge
region with radius $R_{r}$. The BS is assumed to be equipped with
two antennas, one for transmission and one for reception, and each
user in the system is equipped with a single antenna. Number of the
UL and DL users in the cell-center region are $K_{cu}$ and $K_{cd}$,
respectively, where $K_{cu}+K_{cd}=K_{c}$, while number of the UL
and DL users in the cell-edge region are $K_{eu}$ and $K_{ed}$,
respectively, where $K_{eu}+K_{ed}=K_{e}$. The users are uniformly
distributed in the area, where the cell-center users can communicate
directly with the BS, while the cell-edge users transmit and receive
their messages through the STAR-RIS. It is known that the STAR-RIS
is most likely to be installed on the buildings, and hence it can
create channels dominated by the line-of-sight (LoS) path along with
the scatterers. Thus, Rician fading model is considered for the STAR-RIS
related channels. On the other hand, Rayleigh fading is assumed for
the BS to users and user to user channels due to the wealth of scatterers.

\subsection{STAR-RIS Protocol}

In this work, the ES protocol has been implemented at the STAR-RIS.
In the ES scheme, the signal incident upon each SAR-RIS element is
split into transmitted and reflected signals, with ES coefficients
(also named amplitude coefficients) $\rho_{n}^{r}$ and $\rho_{n}^{t}$
($\rho_{n}^{r}+\rho_{n}^{t}\leq1$). Thus, the transmission ($t$)
and reflection ($r$) coefficient matrices of the STAR-RIS can be
written as $\Theta_{k}=\textrm{diag}\left(\rho_{1}^{k}\theta_{1}^{k},....,\rho_{N}^{k}\theta_{N}^{k}\right),k\in\left\{ t,r\right\} $,
where $\theta_{n}^{k}=e^{j\phi_{n}^{k}}$, $\rho_{n}^{k}\in\left[0,1\right]$,
$\rho_{n}^{k}>1$ and $\left|\theta_{n}^{k}\right|=1$.

\section{Data Transmission and Ergodic Rates\label{sec:NOMA-Pairing}}

In the system model under consideration, the FD-BS serves the DL and
UL users by employing NOMA-pairing scheme as follows. In each time
slot, the BS transmits signals to two DL users, a DL cell-center user
(near/strong user), and a DL cell-edge user (far/weak user), and in
the same time slot, the BS receives signals from two UL users, an
UL cell-center user (near/strong user), and an UL cell-edge user (far/weak
user). It is worth mentioning that, the users benefit from NOMA transmissions
over OMA transmission by satisfying the necessary condition that the
achievable rate of NOMA $\left(R_{u_{i}}^{NOMA}\right)$ is larger
than that of OMA, $\left(R_{u_{i}}^{OMA}\right)$. This condition
can be expressed as, $\gamma_{u_{i}}^{NOMA}>\sqrt{1+\gamma_{u_{i}}^{OMA}}-1$,
where $\gamma_{u_{i}}$is the signal to interference and noise ratio
(SINR) at user $i$.

\subsection{DL}

In the DL mode, in a given time slot, the BS transmits the following
superimposed signal

\begin{equation}
s=\stackrel[i=1]{2}{\sum}\sqrt{\alpha_{i}}x_{u_{id}},
\end{equation}

\noindent where $\alpha_{i}$ is the power allocation coefficient
of user $i$ with $\alpha_{1}+\alpha_{2}=1$, and $x_{u_{id}}$ is
the information signal of user $i$ with unit variance.

\subsubsection{Cell-center/strong user}

The received signal at the DL cell-center/strong user (user 1) can
be expressed as

\[
y_{u_{1d}}=\stackrel[i=1]{2}{\sum}\sqrt{P_{b_{i}}}x_{u_{id}}\left(\sqrt{l_{b,u_{1d}}^{-m}}h_{b,u_{1d}}+\right.
\]

\begin{equation}
\left.\sqrt{l_{b,r}^{-m}l_{r,u_{1d}}^{-m}}\mathbf{g}_{r,u_{1d}}\Theta_{t}\mathbf{g}_{b,r}\right)+I_{u_{1d}}+n_{u_{1d}},
\end{equation}

\noindent where $P_{b_{i}}=\alpha_{i}P_{b}$, $P_{b}$ is the BS transmit
power, $\alpha_{1},\alpha_{2}$ are the power coefficients of the
cell-center and cell-edge users, respectively, with $\alpha_{1}<\alpha_{2}$,
$l_{b,u_{1d}}^{-m},l_{b,r}^{-m},l_{r,u_{1d}}^{-m}$ represent the
path-loss between the BS and DL cell-center user, BS and RIS, RIS
and DL cell-center user, respectively, $m$ is the path-loss exponent,
$h_{b,u_{1d}}\thicksim CN(0,1)$ is the channel between the BS and
DL cell-center user, and $\mathbf{g}_{_{r,u_{1d}}}=\left(\sqrt{\frac{\kappa_{r,u_{1d}}}{\kappa_{r,u_{1d}}+1}}\bar{\mathbf{g}}_{_{r,u_{1d}}}+\sqrt{\frac{1}{\kappa_{r,u_{1d}}+1}}\tilde{\mathbf{g}}_{_{r,u_{1d}}}\right)\in C^{1\times N},\mathbf{g}_{b,r}=\left(\sqrt{\frac{\kappa_{b,r}}{\kappa_{b,r}+1}}\bar{\mathbf{g}}_{b,r}+\sqrt{\frac{1}{\kappa_{b,r}+1}}\tilde{\mathbf{g}}_{b,r}\right)\text{\ensuremath{\in}}\mathbb{C}^{N\text{\texttimes}1}$
are the channel vectors between the DL cell-center user and the RIS,
and the BS and the RIS, respectively, $\kappa_{r,u_{1d}}$ and $\kappa_{b,r}$
are the Rician factors, $\bar{\mathbf{g}}_{_{r,u_{1d}}}$ and $\bar{\mathbf{g}}_{b,r}$
are the LoS components and $\tilde{\mathbf{g}}_{_{r,u_{1d}}},$ $\tilde{\mathbf{g}}_{b,r}$
are the non-LoS (NLoS) components, $n_{u_{1d}}$ is the additive white
Gaussian noise (AWGN) at the user, $n_{u_{1d}}\sim CN\left(0,\sigma_{u_{1d}}^{2}\right)$,
and $I_{u_{1d}}$ is the interference caused by the UL users and given
by 

\[
I_{u_{1d}}=\underset{\textrm{interference from UL cell center user }}{\underbrace{\sqrt{p_{u_{1u}}}x_{u_{1u}}\left(\sqrt{l_{u_{1d},u_{1u}}^{-m}}h_{u_{1d},u_{1u}}+\right.}}
\]

\[
\underset{\textrm{interference from UL cell center user }}{\underbrace{\left.\sqrt{l_{r,u_{1u}}^{-m}l_{r,u_{1d}}^{-m}}\mathbf{g}_{_{r,u_{1d}}}\Theta_{t}\mathbf{g}_{_{r,u_{1u}}}\right)}}
\]

\[
+\underset{\textrm{interference from UL cell edge user }}{\underbrace{\sqrt{p_{u_{2u}}l_{r,u_{2u}}^{-m}l_{r,u_{1d}}^{-m}}\mathbf{g}_{r,u_{1d}}\Theta_{t}\mathbf{g}_{_{r,u_{2u}}}x_{u_{2u}}}}
\]

\noindent where $p_{u_{1u}},p_{u_{2u}},x_{u_{1u}},x_{u_{2u}}$ are
the transmit powers and signals of the UL cell-center and cell-edge
users, $h_{u_{1d},u_{1u}}\thicksim CN(0,1)$ is the channel between
the cell-center users in UL and DL modes, and $\mathbf{g}_{_{r,u_{1u}}}=\left(\sqrt{\frac{\kappa_{r,u_{1u}}}{\kappa_{r,u_{1u}}+1}}\bar{\mathbf{g}}_{_{r,u_{1u}}}+\sqrt{\frac{1}{\kappa_{r,u_{1u}}+1}}\tilde{\mathbf{g}}_{_{r,u_{1u}}}\right),\mathbf{g}_{_{r,u_{2u}}}=\left(\sqrt{\frac{\kappa_{r,u_{2u}}}{\kappa_{r,u_{2u}}+1}}\bar{\mathbf{g}}_{_{r,u_{2u}}}+\sqrt{\frac{1}{\kappa_{r,u_{2u}}+1}}\tilde{\mathbf{g}}_{_{r,u_{2u}}}\right)$
are the channel vectors between the UL cell-center user and the RIS,
and the UL cell-edge user and the RIS, respectively. 

The SINR at the DL cell-center user can be written as 

\begin{equation}
\gamma_{u_{1d}}=\frac{P_{b_{1}}A_{u_{1d}}}{\Xi P_{b_{2}}A_{u_{1d}}+p_{u_{1u}}C_{u_{1d}}+p_{u_{2u}}D_{u_{1d}}+\sigma_{u_{1d}}^{2}},
\end{equation}

\noindent where

\noindent $A_{u_{1d}}=$

\noindent $\left|\sqrt{l_{b,u_{1d}}^{-m}}h_{b,u_{1d}}+\sqrt{l_{b,r}^{-m}l_{r,u_{1d}}^{-m}}\mathbf{g}_{_{r,u_{1d}}}\Theta_{t}\mathbf{g}_{b,r}\right|^{2},$

\noindent $C_{u_{1d}}=$

\noindent $\left|\sqrt{l_{u_{1d},u_{1u}}^{-m}}h_{u_{1d},u_{1u}}+\sqrt{l_{r,u_{1u}}^{-m}l_{u_{r,u_{1d}}}^{-m}}\mathbf{g}_{_{r,u1d}}\Theta_{t}\mathbf{g}_{_{r,u1u}}\right|^{2}$,

\noindent $D_{u_{1u}}=l_{r,u_{2u}}^{-m}l_{r,u_{1d}}^{-m}\left|\mathbf{g}_{r,u_{1d}}\Theta_{t}\mathbf{g}_{r,u_{2u}}\right|^{2}$,
and $0\leq\Xi\leq1$ is the fractional error factor that corresponds
to a fraction of the power that remains as interference due to imperfect
SIC. In NOMA pairing scheme, the rate for the strong user to detect
the weak user signal, $R_{u_{1d}\rightarrow u_{2d}}$, should be larger
than or equal to the rate of the weak user $R_{u_{2d}}$. 
\begin{thm}
The ergodic DL rate of the cell-center user in passive STAR-RIS-aided
FD communication systems under Rician fading channels can be calculated
by

\begin{eqnarray*}
\mathcal{E}\left[R_{u_{1d}}\right] & \approx & \log_{2}\left(1+\right.
\end{eqnarray*}

\begin{equation}
\left.\frac{P_{b_{1}}x_{1_{u_{1d}}}}{\Xi P_{b_{2}}x_{1_{u_{1d}}}+p_{u_{1u}}y_{1_{u_{1d}}}+p_{u_{2u}}y_{2_{u_{1d}}}+\sigma_{u_{1d}}^{2}}\right)
\end{equation}

\noindent where

\begin{equation}
x_{1_{u_{1d}}}=\frac{2\left(1+R\right)^{-m}\left(-1+R^{2}+mR\left(1+R\right)+\left(1+R\right)^{m}\right)}{\left(m-2\right)\left(m-1\right)R^{2}}+l_{b,r}^{-m}\Upsilon\left(\varpi_{b,r}^{r,u_{1d}}\xi_{1}+\hat{\varpi}_{b,r}^{r,u_{1d}}\right),\label{eq:5}
\end{equation}

\begin{equation}
y_{1_{u_{1d}}}=\left(\varrho+\Upsilon^{2}\left(\varpi_{r,u1u}^{r,u_{1d}}\xi_{2}+\hat{\varpi}_{r,u1u}^{r,u_{1d}}\right)\right)\label{eq:7}
\end{equation}

\begin{equation}
y_{2_{u_{1d}}}=\frac{2\left(1+R_{r}\right)^{-m}\left(-1+R_{r}^{2}+mR_{r}\left(1+R_{r}\right)+\left(1+R_{r}\right)^{m}\right)\Upsilon\left(\varpi_{r,u2u}^{r,u_{1d}}\xi_{3}+\hat{\varpi}_{r,u2u}^{r,u_{1d}}\right)}{\left(m-2\right)\left(m-1\right)R_{r}^{2}}\label{eq:9}
\end{equation}

\noindent where $\varpi_{y}^{x}=\frac{\kappa_{x}}{\kappa_{x}+1}\frac{\kappa_{y}}{\kappa_{y}+1}$,
$\hat{\varpi}_{y}^{x}=\frac{\kappa_{x}}{\kappa_{x}+1}\frac{\stackrel[n=1]{N}{\sum}\left|\rho_{n}^{k}\right|^{2}}{\kappa_{y}+1}+\frac{\kappa_{y}}{\kappa_{y}+1}\frac{\stackrel[n=1]{N}{\sum}\left|\rho_{n}^{k}\right|^{2}}{\kappa_{x}+1}+\frac{1}{\kappa_{x}+1}\frac{\stackrel[n=1]{N}{\sum}\left|\rho_{n}^{k}\right|^{2}}{\kappa_{y}+1}$
,

\noindent $\Upsilon=\stackrel[j=1]{C}{\sum}\textrm{H}_{j}\,\left(1+\left(R\, r_{j}+R\right)\right)^{-m}\frac{2\left(R\, r_{j}+R\right)}{\pi R^{2}}\cos^{-1}\left(\frac{1}{\left(R\, r_{j}+R\right)}\left(r_{1}+\frac{\left(\left(R\, r_{j}+R\right)^{2}-r_{1}^{2}\right)}{2\left(R+r_{1}\right)}\right)\right)$
,

\noindent $\varrho=\left(\frac{2}{\left(2-3m+m^{2}\right)R^{2}}-\frac{2F\left(\left\{ \frac{1}{2},-1+\frac{m}{2},-\frac{1}{2}+\frac{m}{2}\right\} ,\left\{ \frac{-1}{2},1\right\} ,4R^{2}\right)}{\left(2-3m+m^{2}\right)R^{2}}-F\left(\left\{ \frac{3}{2},\frac{1}{2}+\frac{m}{2},\frac{m}{2}\right\} ,\left\{ \frac{1}{2},3\right\} ,4R^{2}\right)\right.$

\noindent $\left.+\frac{64mR\, F\left(\left\{ 2,\frac{1}{2}+\frac{m}{2},1+\frac{m}{2}\right\} ,\left\{ \frac{3}{2},\frac{7}{2}\right\} ,4R^{2}\right)}{15\pi}-\frac{64mR\, F\left(\left\{ 2,\frac{1}{2}+\frac{m}{2},1+\frac{m}{2}\right\} ,\left\{ \frac{5}{2},\frac{5}{2}\right\} ,4R^{2}\right)}{9\pi}\right)$
and $F\left(.\right)$ is the hypergeometric function .\end{thm}
\begin{IEEEproof}
The proof is provided in Appendix A.
\end{IEEEproof}

\subsubsection{Cell-edge/weak user}

The received signal at the DL cell-edge/weak user (user 2) can be
expressed as

\[
y_{u_{2d}}=\sqrt{P_{b}l_{b,r}^{-m}l_{r,u_{2d}}^{-m}}\mathbf{g}_{_{r,u_{2d}}}\Theta_{r}\mathbf{g}_{b,r}\stackrel[i=1]{2}{\sum}\sqrt{\alpha_{i}}x_{u_{id}}
\]

\begin{equation}
+I_{u_{2d}}+n_{u_{2d}},
\end{equation}

\noindent where $n_{u_{2d}}$ is the AWGN at the user, $n_{u_{2d}}\sim CN\left(0,\sigma_{u_{2d}}^{2}\right)$
and $I_{u_{2d}}$ is the interference caused by the UL users and given
by 

\[
I_{u_{2d}}=\underset{\textrm{interference from cell-center user}}{\underbrace{\sqrt{p_{u_{1u}}l_{r,u_{2d}}^{-m}l_{r,u_{1u}}^{-m}}\mathbf{g}_{_{r,u_{2d}}}\Theta_{r}\mathbf{g}_{r,u_{1u}}x_{u_{1u}}}}
\]

\[
+\underset{\textrm{interference from cell edge user }}{\underbrace{\sqrt{p_{u_{2u}}l_{r,u_{2d}}^{-m}l_{r,u_{2u}}^{-m}}\mathbf{g}_{_{r,u_{2d}}}\Theta_{r}\mathbf{g}_{r,u_{2u}}x_{u_{2u}}}}.
\]

Thus the SINR at the DL cell-edge user can be written as

\begin{equation}
\gamma_{u_{2d}}=\frac{P_{b_{2}}A_{u_{2d}}}{P_{b_{1}}A_{u_{2d}}+p_{u_{1u}}C_{u_{2d}}+p_{u_{2u}}D_{u_{2d}}+\sigma_{u_{2d}}^{2}},
\end{equation}

\noindent where $A_{u_{2d}}=l_{b,r}^{-m}l_{r,u_{2d}}^{-m}\left|\mathbf{g}_{_{r,u_{2d}}}\Theta_{r}\mathbf{g}_{b,r}\right|^{2}$,
$C_{u_{2d}}=l_{r,u_{2d}}^{-m}l_{r,u_{1u}}^{-m}\left|\mathbf{g}_{r,u_{2d}}\Theta_{r}\mathbf{g}_{r,u_{1u}}\right|^{2}$,

\noindent $D_{u_{2u}}=l_{r,u_{2d}}^{-m}l_{r,u_{2u}}^{-m}\left|\mathbf{g}_{_{r,u_{2d}}}\Theta_{r}\mathbf{g}_{r,u_{2u}}\right|^{2}$. 
\begin{thm}
The ergodic DL rate of the cell-edge user in passive STAR-RIS-aided
FD communication systems under Rician fading channels can be calculated
by

\[
\mathcal{E}\left[R_{u_{2d}}\right]\approx\log_{2}\left(1+\right.
\]

\begin{equation}
\left.\frac{P_{b_{2}}x_{1_{u_{2d}}}}{P_{b_{1}}x_{1_{u_{2d}}}+p_{u_{1u}}y_{1_{u_{2d}}}+p_{u_{2u}}y_{2_{u_{2d}}}+\sigma_{u_{2d}}^{2}}\right)
\end{equation}

\noindent where
\end{thm}
\begin{equation}
x_{1_{u_{2d}}}=l_{b,r}^{-m}\left(\frac{2\left(1+R_{r}\right)^{-m}\left(-1+R_{r}^{2}+mR_{r}\left(1+R_{r}\right)+\left(1+R_{r}\right)^{m}\right)}{\left(m-2\right)\left(m-1\right)R_{r}^{2}}\right)\left(\varpi_{b,r}^{r,u_{2d}}\xi_{4}+\hat{\varpi}_{b,r}^{r,u_{2d}}\right)\label{eq:13}
\end{equation}

\begin{equation}
y_{1_{u_{2d}}}=\frac{2\left(1+R_{r}\right)^{-m}\left(-1+R_{r}^{2}+mR_{r}\left(1+R_{r}\right)+\left(1+R_{r}\right)^{m}\right)\Upsilon\left(\varpi_{r,u_{1u}}^{r,u_{2d}}\xi_{5}+\hat{\varpi}_{r,u_{1u}}^{r,u_{2d}}\right)}{\left(m-2\right)\left(m-1\right)R_{r}^{2}}
\end{equation}

\begin{equation}
y_{2_{u_{2d}}}=\left(\frac{2\left(1+R_{r}\right)^{-m}\left(-1+R_{r}^{2}+mR_{r}\left(1+R_{r}\right)+\left(1+R_{r}\right)^{m}\right)}{\left(m-2\right)\left(m-1\right)R_{r}^{2}}\right)^{2}\left(\varpi_{r,u_{2u}}^{r,u_{2d}}\xi_{6}+\hat{\varpi}_{r,u_{2u}}^{r,u_{2d}}\right)\label{eq:17}
\end{equation}

\begin{IEEEproof}
The proof is provided in Appendix B.
\end{IEEEproof}

\subsection{UL}

In the UL mode, the received signal at the BS can be written as 

\[
y_{b}=\sqrt{p_{u_{1u}}}x_{u_{1u}}\left(\sqrt{l_{b,u_{1u}}^{-m}}h_{b,u_{1u}}+\right.
\]

\[
\left.\sqrt{l_{b,r}^{-m}l_{r,u_{1u}}^{-m}}\mathbf{g}_{b,r}\Theta_{t}\mathbf{g}_{r,u_{1u}}\right)+
\]

\begin{equation}
\sqrt{p_{u_{2u}}l_{b,r}^{-m}l_{r,u_{2u}}^{-m}}\mathbf{g}_{b,r}\Theta_{t}\mathbf{g}_{_{r,u_{2u}}}x_{u_{2u}}+I_{b}+n_{b}\label{eq:17-1}
\end{equation}

\noindent where $n_{b}$ is the AWGN at the BS, $n_{b}\sim CN\left(0,\sigma_{b}^{2}\right)$
and $I_{b}$ is the interference term 
\[
I_{b}=\underset{\textrm{self interference }}{\underbrace{\sqrt{P_{b}}h_{b,b}s}}+\underset{\textrm{reflection of the downlink signal}}{\underbrace{\sqrt{P_{b}l_{b,r}^{-m}l_{b,r}^{-m}}\mathbf{g}_{b,r}\Theta_{t}\mathbf{g}_{b,r}s}}.
\]

Several self-interference suppression (SIS) methods have been proposed
in the literature to eliminate the self-interference at the BS \cite{SI1,SI2}.
Applying these interference cancellation techniques can reduce the
self-interference to the background noise level. In this work, similar
to \cite{SI1,SI2}, the residual self-interference is assumed to be
zero-mean Gaussian distributed with variance $V$, $\tilde{s}\sim CN\left(0,V\right)$%
\footnote{According to the central limit theorem, this Gaussian assumption might
occur in practice due to the several sources of imperfection cancellation
stages. Otherwise, the Gaussian assumption can represent the worst-case
or lower-bound of the achievable data rate.%
}. Thus, the interference term $I_{b}$ can be rewritten as 

\[
I_{b}=\underset{\textrm{residual self interference }}{\underbrace{\tilde{s}}}+\underset{\textrm{reflection of the downlink signal}}{\underbrace{\sqrt{P_{b}l_{b,r}^{-m}l_{b,r}^{-m}}\mathbf{g}_{b,r}^{H}\Theta_{t}\mathbf{g}_{b,r}s}}.
\]

Based on the experimental results, the variance of the residual self
interference can be considered mathematically as $V=\beta P_{b}^{\lambda}$,
where the constants $\beta$ and $\lambda$ ($0\le\lambda\le1$) reflect
the efficiency of the cancellation technique.

\subsubsection{Cell-Center user}

Now, we can write the SINR at the BS to detect the UL cell-center/strong
user message as 

\begin{equation}
\gamma_{u_{1u}}=\frac{p_{u_{1u}}A_{u_{1u}}}{p_{u_{2u}}B_{u_{1u}}+P_{b}C_{u_{1u}}+\left|\tilde{s}\right|^{2}+\sigma_{b}^{2}},
\end{equation}

\noindent where

\noindent $A_{u_{1u}}=\left|\sqrt{l_{b,u_{1u}}^{-m}}h_{b,u_{1u}}+\sqrt{l_{b,r}^{-m}l_{r,u_{1u}}^{-m}}\mathbf{g}_{b,r}\Theta_{t}\mathbf{g}_{r,u_{1u}}\right|^{2}$,

\noindent $B_{u_{1u}}=l_{b,r}^{-m}l_{r,u_{2u}}^{-m}\left|\mathbf{g}_{b,r}\Theta_{t}\mathbf{g}_{r,u_{2u}}\right|^{2}$,
$C_{u_{1u}}=l_{b,r}^{-m}l_{b,r}^{-m}\left|\mathbf{g}_{b,r}\Theta_{t}\mathbf{g}_{b,r}^{H}\right|^{2}$.
\begin{thm}
The ergodic UL rate of the cell-center user in passive RIS-aided FD
communication systems under Rician fading channels can be calculated
by

\[
\mathcal{E}\left[R_{u_{1u}}\right]\approx\log_{2}\left(1+\right.
\]

\begin{equation}
\left.\frac{p_{u_{1u}}x_{1_{u_{1u}}}}{p_{u_{2u}}y_{1_{u_{1u}}}+P_{b}y_{2_{u_{1u}}}+V+\sigma_{b}^{2}}\right),\label{eq:19}
\end{equation}

\noindent where
\end{thm}
\begin{equation}
x_{1_{u_{1u}}}=\frac{2\left(1+R\right)^{-m}\left(-1+R^{2}+mR\left(1+R\right)+\left(1+R\right)^{m}\right)}{\left(m-2\right)\left(m-1\right)R^{2}}+l_{b,r}^{-m}\Upsilon\left(\varpi_{b,r}^{r,u_{1u}}\xi_{7}+\hat{\varpi}_{b,r}^{r,u_{1u}}\right)\label{eq:21}
\end{equation}

\begin{equation}
y_{1_{u_{1u}}}=l_{b,r}^{-m}\frac{2\left(1+R_{r}\right)^{-m}\left(-1+R_{r}^{2}+mR_{r}\left(1+R_{r}\right)+\left(1+R_{r}\right)^{m}\right)\left(\varpi_{b,r}^{r,u_{2u}}\xi_{8}+\hat{\varpi}_{b,r}^{r,u_{2u}}\right)}{\left(m-2\right)\left(m-1\right)R_{r}^{2}}
\end{equation}

\[
y_{2_{u_{1u}}}=l_{b,r}^{-m}l_{b,r}^{-m}\left(\left(\frac{\kappa_{b,r}}{\kappa_{b,r}+1}\right)^{2}\xi_{9}+2\frac{\kappa_{b,r}}{\kappa_{b,r}+1}\frac{1}{\kappa_{b,r}+1}\stackrel[n=1]{N}{\sum}\left|\rho_{n}^{k}\right|^{2}+\right.
\]

\begin{equation}
\left.\left(\frac{1}{\kappa_{b,r}+1}\right)^{2}\left(2\stackrel[n=1]{N}{\sum}\left|\rho_{n}^{k}\right|^{2}+\stackrel[n_{1}=1]{N}{\sum}\stackrel[n_{2}\neq n_{1}]{N}{\sum}\left(\rho_{n_{1}}^{k}e^{j\phi_{n_{1}}^{k}}\right)\left(\rho_{n_{2}}^{k}e^{j\phi_{n_{2}}^{k}}\right)^{H}\right)+2\frac{\kappa_{b,r}}{\kappa_{b,r}+1}\frac{1}{\kappa_{b,r}+1}\left(\zeta\stackrel[n=1]{N}{\sum}\rho_{n}^{k}e^{j\phi_{n}^{k}}\right)\right)\label{eq:24}
\end{equation}

\begin{IEEEproof}
The proof is provided in Appendix C
\end{IEEEproof}

\subsubsection{Cell-edge user}

The SINR at the BS to detect the UL cell-edge/weak user message can
be written as

\begin{equation}
\gamma_{u_{2u}}=\frac{p_{u_{2u}}A_{u_{2u}}}{\Xi p_{u_{1u}}B_{u_{2u}}+P_{b}C_{u_{2u}}+\left|\tilde{s}\right|^{2}+\sigma_{b}^{2}},
\end{equation}

\noindent where $A_{u_{2u}}=l_{b,r}^{-m}l_{r,u_{2u}}^{-m}\left|\mathbf{g}_{b,r}\Theta_{t}\mathbf{g}_{r,u_{2u}}\right|^{2}$,

\noindent $B_{u_{2u}}=$

\noindent $\left|\sqrt{l_{b,u_{1u}}^{-m}}h_{b,u_{1u}}+\sqrt{l_{b,r}^{-m}l_{r,u_{1u}}^{-m}}\mathbf{g}_{b,r}\Theta_{t}\mathbf{g}_{r,u_{1u}}\right|^{2}$,

\noindent $C_{u_{2u}}=l_{b,r}^{-m}l_{b,r}^{-m}\left|\mathbf{g}_{b,r}\Theta_{t}\mathbf{g}_{b,r}^{H}\right|^{2}$.
\begin{thm}
The ergodic UL rate of the cell-edge user in passive RIS-aided FD
communication systems under Rician fading channels can be calculated
by

\[
\mathcal{E}\left[R_{u_{2u}}\right]\approx\log_{2}\left(1+\right.
\]

\begin{equation}
\left.\frac{p_{u_{2u}}x_{1_{u_{2u}}}}{\Xi p_{u_{1u}}y_{1_{u_{2u}}}+P_{b}y_{2_{u_{2u}}}+V+\sigma_{b}^{2}}\right)\label{eq:25}
\end{equation}

\noindent where $x_{1_{u_{2u}}}=y_{1_{u1u}}$, $y_{1_{u_{2u}}}=x_{1_{u1u}}$,$y_{2_{u_{2u}}}=y_{2_{u_{1u}}}$.
\end{thm}
\textbf{\emph{Remark.}} All the ergodic rate expressions in this work
are presented in closed form. Thus, the impact of the system parameters
on the total achievable data rates can be observed, and the optimal
design can be attained. Simpler expressions can be easily obtained
by relaxing the practical assumptions, such as perfect SIC, ignoring
the signals reflected from the RIS to the cell center users, assuming
the users are located in fixed locations, and perfect self interference
cancellation. Accordingly, the ergodic rate expressions presented
in Theorems 1-4 can be simplified into, 

\begin{equation}
\mathcal{E}\left[R_{u_{1d}}\right]\approx\log_{2}\left(1+\frac{P_{b_{1}}l_{b,u_{1d}}^{-m}}{p_{u_{1u}}l_{u_{1d},u_{1u}}^{-m}+y_{_{u_{1d}}}+\sigma_{u_{1d}}^{2}}\right)
\end{equation}

\begin{eqnarray}
\mathcal{E}\left[R_{u_{2d}}\right] & \approx & \log_{2}\left(1+\right.\nonumber \\
 &  & \left.\frac{P_{b_{2}}x_{u_{2d}}}{P_{b_{1}}x_{u_{2d}}+y_{2_{u_{2d}}}+y_{3_{u_{2d}}}+\sigma_{u_{2d}}^{2}}\right)
\end{eqnarray}

\begin{equation}
\mathcal{E}\left[R_{u_{1u}}\right]\approx\log_{2}\left(1+\frac{p_{u_{1u}}l_{b,u_{1u}}^{-m}}{y_{2_{u_{1u}}}+\sigma_{b}^{2}}\right)
\end{equation}

\begin{equation}
\mathcal{E}\left[R_{u_{2u}}\right]\approx\log_{2}\left(1+\frac{x_{u_{2u}}}{\sigma_{b}^{2}\left(\kappa_{r,u2u}+1\right)\left(\kappa_{b,r}+1\right)}\right)
\end{equation}

\noindent where 

\noindent $y_{_{u_{1d}}}=p_{u_{2u}}l_{r,u_{2u}}^{-m}l_{r,u_{1d}}^{-m}\left(\varpi_{r,u_{2u}}^{r,u_{1d}}\xi_{3}+\hat{\varpi}_{r,u_{2u}}^{r,u_{1d}}\right)$,

\noindent $x_{u_{2d}}=l_{b,r}^{-m}l_{r,u_{2d}}^{-m}\left(\varpi_{b,r}^{r,u_{2d}}\xi_{4}+\hat{\varpi}_{b,r}^{r,u_{2d}}\right)$

\noindent $y_{2_{u_{2d}}}=p_{u_{1u}}l_{r,u_{2d}}^{-m}l_{r,u_{1u}}^{-m}\left(\varpi_{u_{1u,r}}^{r,u_{2d}}\xi_{5}+\hat{\varpi}_{u_{1u,r}}^{r,u_{2d}}\right)$,

\noindent $y_{3_{u_{2d}}}=p_{u_{2u}}l_{r,u_{2d}}^{-m}l_{r,u_{2u}}^{-m}\left(\varpi_{r,u_{2u}}^{r,u_{2d}}\xi_{6}+\hat{\varpi}_{r,u_{2u}}^{r,u_{2d}}\right)$

\noindent $y_{2_{u_{1u}}}=p_{u_{2u}}l_{b,r}^{-m}l_{r,u_{2u}}^{-m}\left(\varpi_{b,r}^{r,u_{2u}}\xi_{8}+\hat{\varpi}_{b,r}^{r,u_{2u}}\right)$

\noindent $x_{u_{2u}}=p_{u_{2u}}l_{b,r}^{-m}l_{r,u_{2u}}^{-m}\kappa_{r,u2u}\kappa_{b,r}\xi_{8}+$

\noindent $p_{u_{2u}}l_{b,r}^{-m}l_{r,u_{2u}}^{-m}\stackrel[n=1]{N}{\sum}\left|\rho_{n}^{k}\right|^{2}\left(\kappa_{r,u2u}+\kappa_{b,r}+1\right)$.

\section{Bidirectional Communication\label{sec:Bidirectional-Communication}}

Besides other applications, the above results can be used to analyze
the performance of STAR-RIS networks with bidirectional communication
tasks, which can be interpreted as bidirectional relaying where the
BS plays the role of the relay node \cite{bidirectional2}. In this
context, we will focus on the most difficult scenario, when the UL
cell-center user communicates with the DL cell-edge, and the UL cell-edge
user communicates with the DL cell-center user through the FD-BS.
Particularly, in each time slot the UL users transmit their data to
the FD-BS, and the BS broadcasts the data to the DL users using NOMA
paring scheme.

\subsection{UL Cell-edge user to DL Cell-center user}

The received signal at the DL cell-center user can be expressed by

\[
y_{u_{1d}}=\sqrt{p_{u_{2u}}l_{r,u_{2u}}^{-m}l_{r,u_{1d}}^{-m}}\mathbf{g}_{r,u_{1d}}\Theta_{t}\mathbf{g}_{_{r,u_{2u}}}x_{u_{2u}}+
\]

\[
\stackrel[i=1]{2}{\sum}\sqrt{P_{b_{i}}}\hat{x}_{u_{iu}}\left(\sqrt{l_{b,u_{1d}}^{-m}}h_{b,u_{1d}}+\right.
\]

\[
\left.\sqrt{l_{b,r}^{-m}l_{r,u_{1d}}^{-m}}\mathbf{g}_{r,u_{1d}}\Theta_{t}\mathbf{g}_{b,r}\right)
\]

\begin{equation}
+I_{u_{1d}}+n_{u_{1d}},
\end{equation}

\noindent where the interference term is given by

\noindent $I_{u_{1d}}=\sqrt{p_{u_{1u}}}x_{u_{1u}}\left(\sqrt{l_{u_{1d},u_{1u}}^{-m}}h_{u_{1d},u_{1u}}+\right.$

$\left.\sqrt{l_{r,u_{1u}}^{-m}l_{r,u_{1d}}^{-m}}\mathbf{g}_{_{r,u_{1d}}}\Theta_{t}\mathbf{g}_{_{r,u_{1u}}}\right)$.
Recall that $\hat{x}_{u_{2u}}=x_{u_{2u}}\left(\bar{i}-\hat{\tau}\right)$,
where $\bar{i}$ denotes the $\bar{i}$th time slot and $\hat{\tau}$
is the decoding time of the message at the BS. Assuming that the processing
delay caused by the detection process at the BS is small compared
to the duration of one channel use, i.e.,$\hat{\tau}\leq\hat{t}_{i+1}\text{\textminus}\hat{t}_{i}$,
where $\hat{t}_{i+1}$ and $\hat{t}_{i}$ denote the $(\bar{i}+1)$th
and the $\bar{i}$th time slots, respectively. Thus, the cell-center
user receives the message $x_{u_{2u}}$ from the BS and the cell-edge
user at approximately the same channel use. Hence, the user can successfully
combine the signal $\hat{x}_{u_{2u}}$ transmitted from the BS and
the signal $x_{u_{2u}}$ from the UL cell-edge user by a proper diversity-combining
technique such as maximum ratio combining (MRC) \cite{mrc}. Consequently,
the achievable rate of the DL cell-center user can be expressed by

\[
R_{u_{c}}=\log_{2}\left(1+\frac{p_{u_{2u}}a_{u_{1d}}}{p_{u_{1u}}C_{u_{1d}}+\sigma_{u_{1d}}^{2}}\right.
\]

\begin{equation}
\left.+\frac{P_{b_{1}}A_{u_{1d}}}{\Xi P_{b_{2}}A_{u_{1d}}+p_{u_{1u}}C_{u_{1d}}+\sigma_{u_{1d}}^{2}}\right),
\end{equation}

\noindent where $a_{u_{1d}}=l_{r,u_{2u}}^{-m}l_{r,u_{1d}}^{-m}\left|\mathbf{g}_{r,u_{1d}}\Theta_{t}\mathbf{g}_{r,u_{2u}}\right|^{2}$.
\begin{thm}
The ergodic rate of the DL cell-center user in passive STAR-RIS-aided
FD bidirectional communication under Rician fading channels can be
calculated by

\begin{equation}
\bar{R}_{c}=\min\left(\bar{R}_{u_{2u}},\bar{R}_{u_{c}}\right)
\end{equation}

where

\[
\bar{R}_{u_{c}}=\log_{2}\left(1+\frac{p_{u_{2u}}y_{2_{u_{1d}}}}{p_{u_{1u}}y_{1_{u_{1d}}}+\sigma_{u_{1d}}^{2}}\right.
\]

\begin{equation}
\left.+\frac{P_{b_{1}}x_{1_{u_{1d}}}}{\Xi P_{b_{2}}x_{1_{u_{1d}}}+p_{u_{1u}}y_{1_{u_{1d}}}+\sigma_{u_{1d}}^{2}}\right),
\end{equation}

\noindent and $\bar{R}_{u_{2u}}=\mathcal{E}\left[R_{u_{2u}}\right]$
is given in (\ref{eq:25}), $x_{1_{u_{1d}}}$,$y_{1_{u_{1d}}}$,$y_{2_{u_{1d}}}$
are defined in (\ref{eq:5})-(\ref{eq:9}).\end{thm}
\begin{IEEEproof}
The proof is based on the analysis in Appendix A.
\end{IEEEproof}

\subsection{UL Cell-center user to DL Cell-edge user}

The received signal at the cell edge user is given by

\[
y_{u_{2d}}=\sqrt{p_{u_{1u}}l_{r,u_{2d}}^{-m}l_{r,u_{1u}}^{-m}}\mathbf{g}_{_{r,u_{2d}}}\Theta_{r}\mathbf{g}_{r,u_{1u}}x_{u_{1u}}
\]

\begin{equation}
+\sqrt{P_{b}l_{b,r}^{-m}l_{r,u_{2d}}^{-m}}\mathbf{g}_{_{r,u_{2d}}}\Theta_{r}\mathbf{g}_{b,r}\stackrel[i=1]{2}{\sum}\sqrt{\alpha_{i}}\hat{x}_{u_{iu}}+I_{u_{2d}}+n_{u_{2d}},
\end{equation}

\noindent where $I_{u_{2d}}=\sqrt{p_{u_{2u}}l_{r,u_{2d}}^{-m}l_{r,u_{2u}}^{-m}}\mathbf{g}_{_{r,u_{2d}}}\Theta_{r}\mathbf{g}_{r,u_{2u}}x_{u_{2u}}$.
Similarly, the achievable rate of the DL cell edge user can be expressed
by

\[
R_{u_{e}}=\log_{2}\left(1+\frac{p_{u_{1u}}a_{u_{2d}}}{D_{u_{2d}}+\sigma_{u_{2d}}^{2}}\right.
\]

\begin{equation}
\left.+\frac{P_{b_{2}}A_{u_{2d}}}{P_{b_{1}}A_{u_{2d}}+p_{u_{2u}}D_{u_{2d}}+\sigma_{u_{2d}}^{2}}\right),
\end{equation}

\noindent where $a_{u_{2d}}=l_{r,u_{2d}}^{-m}l_{r,u_{1u}}^{-m}\left|\mathbf{g}_{r,u_{2d}}\Theta_{r}\mathbf{g}_{r,u_{1u}}\right|^{2}$.
\begin{thm}
The ergodic rate of the DL cell-edge user in passive STAR-RIS-aided
FD bidirectional communication under Rician fading channels can be
calculated by

\begin{equation}
\bar{R}_{e}=\min\left(\bar{R}_{u_{1u}},\bar{R}_{u_{e}}\right)
\end{equation}

where

\[
\bar{R}_{u_{e}}\approx\log_{2}\left(1+\frac{p_{u_{1u}}y_{1_{u_{2d}}}}{p_{u_{2u}}y_{2_{u_{2d}}}+\sigma_{u_{2d}}^{2}}\right.
\]

\begin{equation}
\left.+\frac{P_{b_{2}}x_{1_{u_{2d}}}}{P_{b_{1}}x_{1_{u_{2d}}}+p_{u_{2u}}y_{2_{u_{2d}}}+\sigma_{u_{2d}}^{2}}\right),
\end{equation}

\noindent and $\bar{R}_{u_{1u}}=\mathcal{E}\left[R_{u_{1u}}\right]$
is given in (\ref{eq:19}), $x_{1_{u_{2d}}},y_{1_{u_{2d}}},y_{2_{u_{2d}}}$are
defined in (\ref{eq:13})-(\ref{eq:17}).\end{thm}
\begin{IEEEproof}
The proof is based on the analysis in Appendix B.
\end{IEEEproof}

\section{Systems Design\label{sec:System-Design}}

As we can notice from the previous sections, the ergodic rates are
function of the STAR-RIS amplitudes and phase shifts and the UL and
DL transmit powers. Therefore, these parameters can be optimized to
maximize the total weighted sum rate. Accordingly, the optimization
problem for both scenarios can be formulated as 

\begin{eqnarray}
 & \underset{\rho,\mathbf{\theta},\mathbf{p}}{\max}\, f\left(\theta,\rho,\mathbf{p}\right)\nonumber \\
(C.1) & \textrm{s.t}\:\stackrel[i=1]{2}{\sum}p_{u_{iu}}+P_{b_{i}}\leq P_{t},\, p_{u_{iu}}\geq0,P_{b_{i}}\geq0\nonumber \\
(C.2) & R_{u_{1d}\rightarrow u_{2d}}\geqslant R_{u_{2d}}\nonumber \\
(C.3) & R_{u_{2d}}\geq R_{d_{th}},R_{u_{2u}}\geq R_{u_{th}}\nonumber \\
(C.4) & \left(\rho_{n}^{r}\right)+\left(\rho_{n}^{t}\right)=1,\forall n\in N\nonumber \\
(C.5) & \rho_{n}^{k}\geq0,\left|\theta_{n}^{k}\right|=1,\forall n\in N\label{eq:29-1}
\end{eqnarray}

\noindent where $f\left(\theta,\rho,\mathbf{p}\right)=\stackrel[i=1]{2}{\sum}\left(\omega_{id}\bar{R}_{u_{id}}+\omega_{iu}\bar{R}_{u_{iu}}\right)$,
and $f\left(\theta,\rho,\mathbf{p}\right)=\omega_{u_{c}}\bar{R}_{u_{c}}+\omega_{u_{e}}\bar{R}_{u_{e}}$
for the bidirectional communication scenario, $\mathbf{\mathbf{\theta}}=\left[\theta^{t},\theta^{r}\right]$,
$\mathbf{\mathbf{\mathbf{\rho}}}=\left[\rho^{t},\rho^{r}\right]$,
$\mathbf{p}=\left[p_{u_{1}},p_{u_{2}},P_{b_{1}},P_{b_{2}}\right]^{T}$,
and $P_{t}$ is the total transmission power $P_{t}=\tau P_{b}+\left(1-\tau\right)P_{u}$,
$P_{u}=p_{u_{1}}+p_{u_{2}}$, $0<\tau\leq1$, $\omega_{i}$ are the
weighting factors, which signify the priority assigned to each user.
The first constraint $(C.1)$ upper bounds the network transmit power,
where the transmit powers are larger than zero, while the second constraint
$(C.2)$ provides the fundamental condition for the implementation
of SIC%
\footnote{In UL the BS is the receiver, thus it can achieve SIC in any users
order. %
}. The third constraint $(C.3)$ is required to provide a fair power
allocation scheme for the cell-edge users. The last two constraints
$(C.4)$ and $(C.5)$ for the amplitude and phase shift on each STAR-RIS
element. It is extremely difficult to find a solution for the problem
due to its non convexity in nature. However, the main problem can
be divided into two sub-problems: 1) Fix $\mathbf{p}$ and optimize
$\rho,\mathbf{\theta}$ to maximize the sum rate, 2) Fix $\rho,\mathbf{\theta}$
and reformulate the problem to optimize $\mathbf{p}$.

\subsection{Simultaneous Amplitudes And Phase-Shifts Optimization}

For a given power transmission, the main problem can be simplified
into a sub-problem to maximize the sum rate with the amplitudes and
phase-shifts as follows

\begin{eqnarray}
 & \underset{\rho,\mathbf{\theta}}{\max}\, f\left(\theta,\rho\right)\nonumber \\
 & \textrm{s.t}\:\left(\rho_{n}^{r}\right)+\left(\rho_{n}^{t}\right)=1,\forall n\in N,\nonumber \\
 & \rho_{n}^{k}\geq0,\left|\theta_{n}^{k}\right|=1,\forall n\in N.\label{eq:29}
\end{eqnarray}

\noindent The two optimization variables in (\ref{eq:29}) with the
non-convexity make the problem hard to solve. However, the projected
gradient ascent method (PGAM), can be applied to obtain the optimal
amplitudes and the phase shifts simultaneously. To apply the PGAM,
we denote, $\Phi=\left\{ \theta^{t}\in\mathbb{C}^{N\times1},\theta^{r}\in\mathbb{C}^{N\times1}\left|\left|\theta^{r}\right|=\left|\theta^{t}\right|=1\right.\right\} $,
and 

\noindent $Q=\left\{ \rho^{t},\rho^{r}\in\mathbb{C}^{N\times1}\left|\left(\rho_{i}^{t}\right)+\left(\rho_{i}^{r}\right)=1,\rho_{i}^{k}\geq0\right.\right\} $. 

\noindent Then, we compute the gradients of $f\left(\theta,\rho\right)$
with with respect to $\theta$, $\nabla_{\theta}f\left(\theta^{i},\rho^{i}\right)$,
and $\rho$, $\nabla_{\rho}f\left(\theta^{i},\rho^{i}\right)$. Next
the RIS phases and amplitudes are updated at each iteration using
the following expressions, $\theta^{i+1}=\left(\theta^{i}+\mu_{i}\nabla_{\theta}f\left(\theta^{i},\rho^{i}\right)\right)$
and $\rho^{i+1}=\left(\rho^{i}+\alpha\mu_{i}\nabla_{\rho}f\left(\theta^{i},\rho^{i}\right)\right)$,
where $\mu_{i}$ and $\alpha\mu_{i}$ correspond to the step sizes.
Then we project them onto $\Phi$ and $Q$. The overall algorithm
is summarized in Algorithm 1. 

\noindent 
\begin{algorithm}[H]
Input: Given a tolerance $\epsilon>0$ and the maximum number of iterations
$L$. Initialize $\mathbf{\rho}^{(0)},\textrm{ and }\mathbf{\theta}^{(0)}$
. Set step size $\mu=0.5$.

for $i=0\textrm{ to }L$ do

Evaluate: $f\left(\theta^{(i)},\rho^{(i)}\right)$, then $\nabla_{\theta}f\left(\theta^{(i)},\rho^{(i)}\right)$,
and $\nabla_{\rho}f\left(\theta^{(i)},\rho^{(i)}\right)$

Update: $\theta^{i+1}=\left(\theta^{i}+\mu\nabla_{\theta}f\left(\theta^{i},\rho^{i}\right)\right)$
and $\rho^{i+1}=\left(\rho^{i}+\alpha\mu\nabla_{\rho}f\left(\theta^{i},\rho^{i}\right)\right)$

Evaluate: $f\left(\theta^{(i+1)},\rho^{(i+1)}\right)$

Until Convergence: $f\left(\theta^{(i+1)},\rho^{(i+1)}\right)-f\left(\theta^{(i)},\rho^{(i)}\right)<\epsilon$.

\protect\caption{Optimization Algorithm.}
\end{algorithm}

\subsubsection{\noindent Sub-optimal Design}

\noindent The phase shifts and amplitude coefficients can be obtained
based on the ergodic rate expressions and the cell-edge users channels
as follows. The cell-edge users have higher preference in the STAR-RIS
design, thus as an efficient simple design, the phase shifts of the
elements, $\theta^{r}$ and $\theta^{t}$, can be aligned to the cell-edge
users channels. Hence, the phase shifts can be presented in terms
of the BS-RIS and RIS-user channels as, $\phi_{n}^{k}=-2\pi\frac{d}{\lambda}\left(x_{n}t_{b,u_{2k}}+y_{n}l_{b,u_{2k}}\right),k\in\left\{ t,r\right\} $
where $t_{u_{2k}}=\sin\varphi_{b,r}^{a}\sin\varphi_{b,r}^{e}-\sin\varphi_{r,u_{2k}}^{a}\sin\varphi_{r,u_{2k}}^{e},$
and $l_{u_{2k}}=\cos\varphi_{b,r}^{e}-\cos\varphi_{r,u_{2k}}^{e}$,
while $\varphi_{x,y}^{a},\varphi_{x,y}^{e}$ denote the azimuth and
elevation angles of arrival (AoA) from node $x$ to node $y$, $\lambda$
is the wavelength, $d$ is the elements spacing, and $x_{n}=\left(n-1\right)\textrm{mod}\sqrt{N}$,
$y_{n}=\frac{n-1}{\sqrt{N}}$. Then, the amplitude coefficients $\rho_{n}^{r}$
and $\rho_{n}^{t}$ can be calculated from the ergodic rate expressions
in Theorems 2 and 4 according to the required data rates at cell-edge
users.

\noindent For the bidirectional communication scenario, the total
SINR is a combination of two parts, $\gamma=\gamma_{1}+\gamma_{2}$,
where $\gamma_{1}$ is the SINR due to user-RIS-user link, and $\gamma_{2}$
is the SINR for BS-RIS-user link. As a sub-optimal design, the STAR-RIS
can be designed based on the maximum SINR, i.e., $\max\left(\gamma_{1},\gamma_{2}\right)$.
In case $\gamma_{1}$ is the maximum SINR the phase shifts can be
aligned to the user-RIS-user channels as, $\phi_{n}^{k}=-2\pi\frac{d}{\lambda}\left(x_{n}t_{u_{1\bar{k}},u_{2k}}+y_{n}l_{u_{1\bar{k}},u_{2k}}\right)$,
and in case $\gamma_{2}$ is the maximum SINR, the phase shifts can
be aligned to the BS-RIS-user channels as, $\phi_{n}^{k}=-2\pi\frac{d}{\lambda}\left(x_{n}t_{b,u_{2k}}+y_{n}l_{b,u_{2k}}\right)$.

\subsection{Power Allocation}

The optimal values of the transmission powers can be obtained by reformulating
the problem in (\ref{eq:29}) as 

\begin{eqnarray}
 & \underset{\mathbf{p}}{\max}\,\stackrel[i=1]{2}{\sum}f\left(\mathbf{p}\right)\nonumber \\
 & \textrm{s.t}\:\stackrel[i=1]{2}{\sum}p_{u_{iu}}+P_{b_{i}}\leq P_{t},\, p_{u_{iu}}\geq0,P_{b_{i}}\geq0\nonumber \\
 & \bar{R}_{u_{1d}\rightarrow u_{2d}}\geqslant\bar{R}_{u_{2d}}\nonumber \\
 & \bar{R}_{u_{2d}}\geq R_{d_{th}},\bar{R}_{u_{2u}}\geq R_{u_{th}}.
\end{eqnarray}

\noindent This problem can be solved optimally by applying several
schemes such as monotonic optimization, and block coordinate descent
(BCD) iterative algorithms as explained in \cite{powerallocation1,powerallocation2},
the details are omitted here due to paucity of space. Additionally
in this work, based on the derived ergodic rate expressions we present
a simple and efficient power allocation scheme.

\subsubsection{\noindent Sub-optimal Design}

\noindent The DL and UL power transmission can be obtained by solving
simple equations as follows. The transmission power for the cell-edge
users $p_{u_{2u}}$ and $P_{b_{2}}$ can be obtained by satisfying
the last fairness constraint with equality, e.g., $\bar{R}_{u_{2d}}=R_{d_{th}},\bar{R}_{u_{2u}}=R_{u_{th}}$
where $\bar{R}_{u_{2d}}$ and $\bar{R}_{u_{2u}}$ are presented in
Theorems 2 and 4, respectively. Also, $P_{b_{1}}$can be found by
achieving the required SIC condition, $\bar{R}_{u_{1d}\rightarrow u_{2d}}=R_{d_{th}}$,
and hence $p_{u_{1u}}=P_{t}-\left(p_{u_{2u}}+\stackrel[i=1]{2}{\sum}P_{b_{i}}\right)$.
Considering these equations together, we can find the DL and UL power
values as 

\begin{eqnarray}
P_{b_{1}} & = & P_{t}\frac{\left(c_{1}-c_{2}\right)}{\left(b_{2}-b_{1}\right)}-\frac{\left(q_{1}-q_{2}\right)}{\left(b_{2}-b_{1}\right)}\\
P_{b_{2}} & = & P_{t}\left(b_{1}\frac{\left(c_{1}-c_{2}\right)}{\left(b_{2}-b_{1}\right)}+b_{2}\right)-b_{1}\frac{\left(q_{1}-q_{2}\right)}{\left(b_{2}-b_{1}\right)}-q_{2}
\end{eqnarray}

\begin{eqnarray}
p_{u_{2u}} & = & P_{t}\left(\frac{\left(c_{1}-c_{2}\right)}{\left(b_{2}-b_{1}\right)}R_{u_{th}}\frac{y_{2_{u_{2u}}}}{x_{1_{u_{2u}}}}+\right.\nonumber \\
 &  & \left.\left(b_{2}\frac{\left(c_{1}-c_{2}\right)}{\left(b_{2}-b_{1}\right)}+c_{2}\right)R_{u_{th}}\frac{y_{2_{u_{2u}}}}{x_{1_{u_{2u}}}}\right)+\hat{c}
\end{eqnarray}

\noindent \begin{flushleft}
where $b_{i}=\frac{a_{i}-e_{i}R_{d_{th}}-e_{i}R_{d_{th}}R_{u_{th}}\frac{y_{2_{u_{2u}}}}{x_{1_{u_{2u}}}}}{1+e_{i}R_{d_{th}}+e_{i}R_{d_{th}}R_{u_{th}}\frac{y_{2_{u_{2u}}}}{x_{1_{u_{2u}}}}}$,
$c_{i}=\frac{e_{i}R_{d_{th}}}{1+e_{i}R_{d_{th}}+e_{i}R_{d_{th}}R_{u_{th}}\frac{y_{2_{u_{2u}}}}{x_{1_{u_{2u}}}}}$,
$q_{i}=\frac{w_{i}}{1+e_{i}R_{d_{th}}+e_{i}R_{d_{th}}R_{u_{th}}\frac{y_{2_{u_{2u}}}}{x_{1_{u_{2u}}}}}$,
$a_{i}=\frac{R_{d_{th}}+R_{u_{th}}R_{d_{th}}\frac{y_{2_{u_{id}}}}{x_{1_{u_{id}}}}\frac{y_{2_{u_{2u}}}}{x_{1_{u_{2u}}}}}{1-R_{u_{th}}R_{d_{th}}\frac{y_{2_{u_{id}}}}{x_{1_{u_{id}}}}\frac{y_{2_{u_{2u}}}}{x_{1_{u_{2u}}}}}$,
$e_{i}=\frac{y_{1_{u_{id}}}}{x_{1_{u_{id}}}\left(1-R_{u_{th}}R_{d_{th}}\frac{y_{2_{u_{id}}}}{x_{1_{u_{id}}}}\frac{y_{2_{u_{2u}}}}{x_{1_{u_{2u}}}}\right)}$,
$w_{i}=a_{i}R_{d_{th}}R_{u_{th}}\frac{V}{x_{1_{u_{2u}}}}-e_{i}R_{d_{th}}R_{u_{th}}\frac{\sigma_{b}^{2}}{x_{1_{u_{2u}}}}-z_{i}$,
$z_{i}=\frac{\eta_{i}}{1-R_{u_{th}}R_{d_{th}}\frac{y_{2_{u_{id}}}}{x_{1_{u_{id}}}}\frac{y_{2_{u_{2u}}}}{x_{1_{u_{2u}}}}}$,
$\eta_{i}=R_{u_{th}}R_{d_{th}}\frac{y_{2_{u_{id}}}}{x_{1_{u_{id}}}}\frac{V}{x_{1_{u_{2u}}}}+R_{u_{th}}R_{d_{th}}\frac{y_{2_{u_{id}}}}{x_{1_{u_{id}}}}\frac{\sigma_{b}^{2}}{x_{1_{u_{2u}}}}+R_{d_{th}}\frac{\sigma_{u_{2d}}^{2}}{x_{1_{u_{2d}}}}$
and $\hat{c}=-\frac{\left(q_{1}-q_{2}\right)}{\left(b_{2}-b_{1}\right)}R_{u_{th}}\frac{y_{2_{u_{2u}}}}{x_{1_{u_{2u}}}}-b_{2}\frac{\left(q_{1}-q_{2}\right)}{\left(b_{2}-b_{1}\right)}R_{u_{th}}\frac{y_{2_{u_{2u}}}}{x_{1_{u_{2u}}}}-q_{2}R_{u_{th}}\frac{y_{2_{u_{2u}}}}{x_{1_{u_{2u}}}}+R_{u_{th}}\frac{V}{x_{1_{u_{2u}}}}+R_{u_{th}}\frac{\sigma_{b}^{2}}{x_{1_{u_{2u}}}}$.
\par\end{flushleft}

\noindent \begin{flushleft}
 
\par\end{flushleft}
\begin{IEEEproof}
The proof is provided in Appendix D.
\end{IEEEproof}

\section{Numerical Results\label{sec:Numerical-Results}}

In this section, we present simulation and numerical results to validate
our analysis, and to examine the effects of different parameters on
the overall system performance. The radius of the cell-center region
is $R=50m$, while the radius of the cell-edge region is $R_{r}=30m$.
For simplicity, the transmit SNR is defined as $\bar{\gamma}=\frac{P_{t}}{\sigma^{2}}$,
the Rician factors are 3, $\beta=0.001,\lambda=0.1$, $\omega=0.8$
and the path loss exponent is $m=2.7$ \cite{RefMain}. Number of
RIS elements $N=20$, number of users is $K_{c}$=$K_{e}=6$ .

\begin{figure}[H]
\noindent \begin{centering}
\subfloat[\label{fig:2a}Ergodic rates versus transmit SNR$\bar{\gamma}$, with
optimal phase shifts. ]{\noindent \begin{centering}
\includegraphics[scale=0.5]{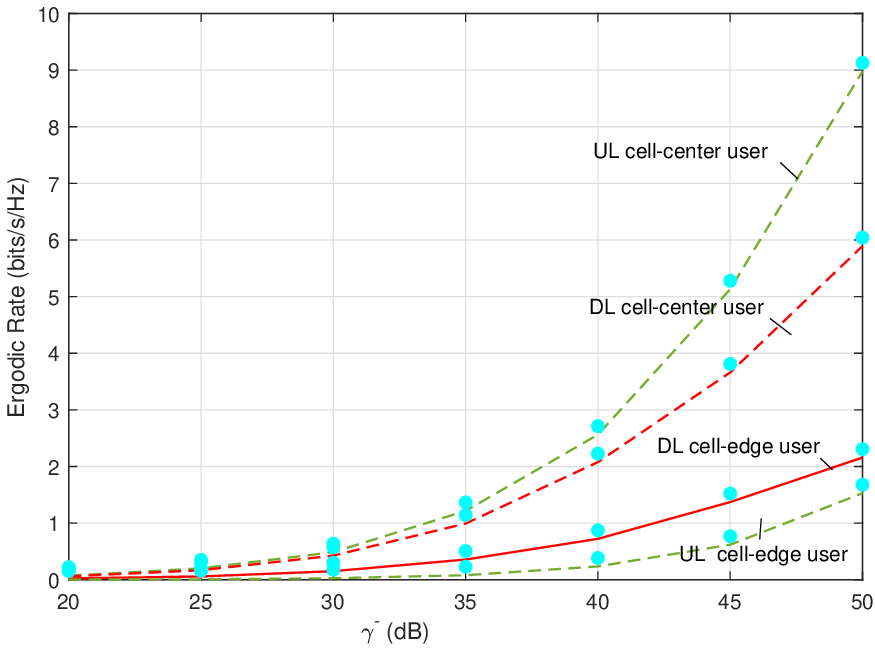}
\par\end{centering}

}\subfloat[\label{fig:2b}Ergodic rates versus transmit SNR$\bar{\gamma}$, with
random phase shifts.]{\noindent \begin{centering}
\includegraphics[scale=0.5]{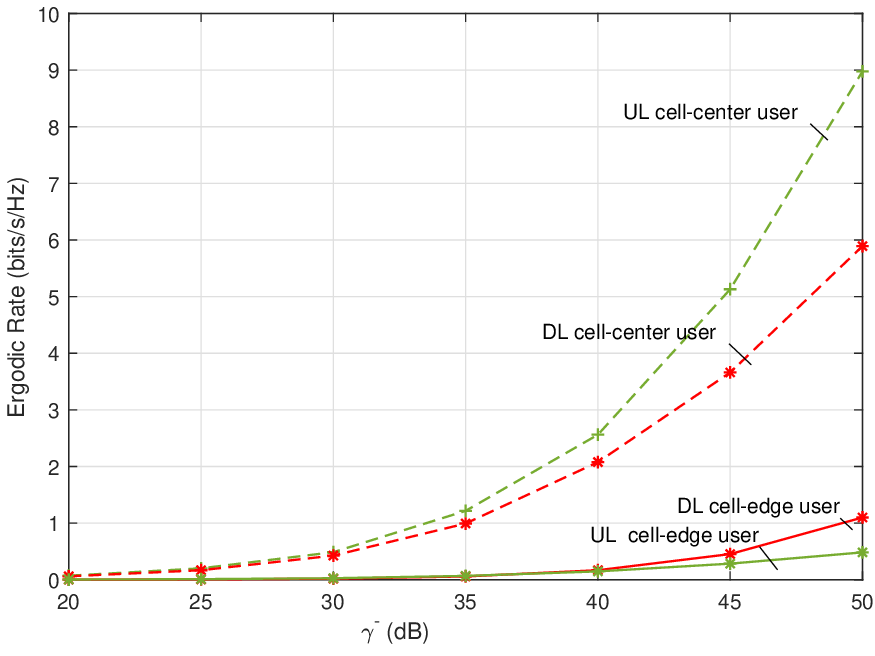}
\par\end{centering}

}
\par\end{centering}

\protect\caption{\label{fig:2}Ergodic rates versus transmit SNR,$\bar{\gamma}$, for
different phase shifts.}

\end{figure}

Firstly, in Fig. \ref{fig:2} we illustrate the ergodic rates versus
the transmit SNR, $\bar{\gamma}$. Fig. \ref{fig:2a} shows the achievable
rates using optimal phase shifts, while the achievable rates with
random phase shifts are presented in Fig. \ref{fig:2b}. It is evident
from these figures that the analytical and simulation results are
in good agreement, which confirms the accuracy of the analysis. It
is also clear that the ergodic rates are enhanced with an increase
in the transmit SNR for both scenarios. In addition, it can be observed
that the ergodic rates of the edge users utilizing the optimal phase
shifts are higher than those utilizing random phase shifts. On the
other hand, the performance of the cell-center users does not depend
essentially on the STAR-RIS phase shifts. It has been also noted that
80\% of the total power has been allocated for the DL transmission,
while 20\% for the UL. This is due to the impact of the interference
power at the DL users. 

\begin{figure}[H]
\noindent \begin{centering}
\subfloat[\label{fig:3a}Ergodic rates versus transmit SNR $\bar{\gamma}$ when
SIC error $\Xi=0.01$ .]{\noindent \begin{centering}
\includegraphics[scale=0.5]{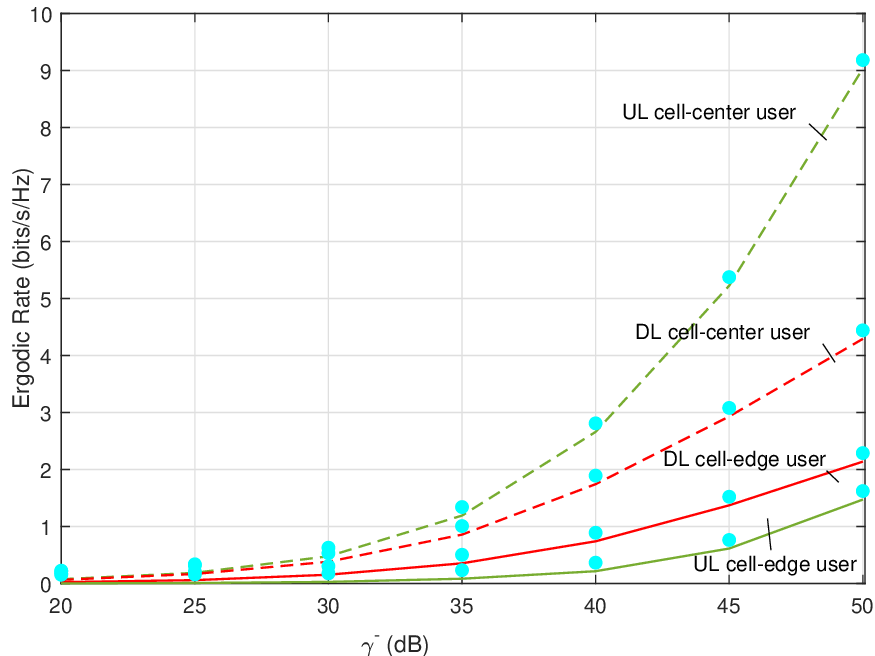}
\par\end{centering}

}\subfloat[\label{fig:3b}Ergodic rates versus transmit SNR $\bar{\gamma}$ when
$\beta=1,\lambda=0.1$. ]{\noindent \begin{centering}
\includegraphics[scale=0.5]{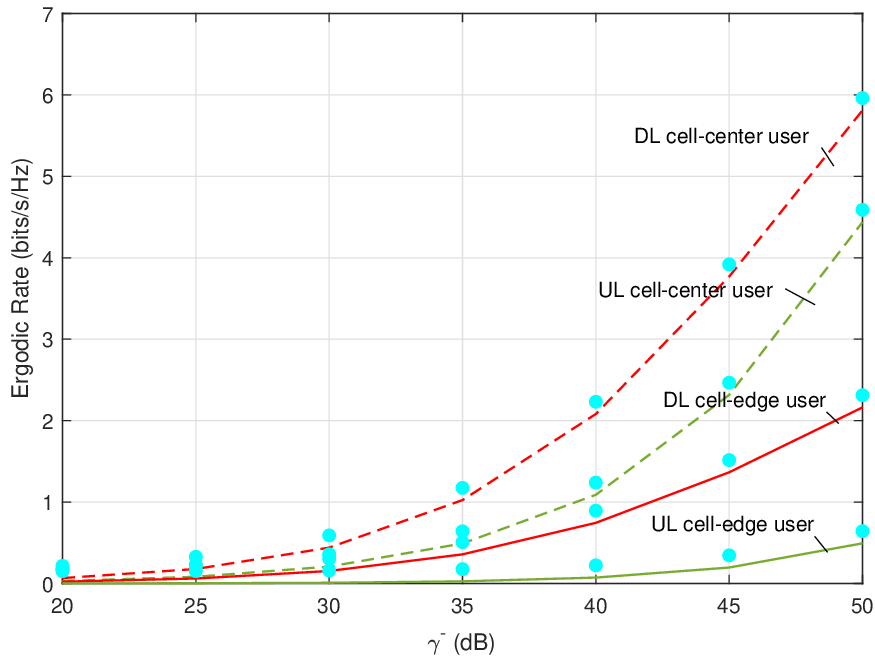}
\par\end{centering}

}
\par\end{centering}

\protect\caption{\label{fig:3}Ergodic rates versus transmit SNR $\bar{\gamma}$ with
SIC error and self interference.}

\end{figure}

To demonstrate the impact of the system impairments on the users performance,
in Fig. \ref{fig:3} we plot the ergodic rates versus the transmit
SNR, $\bar{\gamma}$, for different values of the SIC error factor,
and the variance of the residual self interference. Specifically,
Fig. \ref{fig:3a} depicts the impact of the imperfect SIC on the
achievable rates, while Fig. \ref{fig:3b} investigates the impact
of imperfect self interference suppression, SIS, at the BS. Comparing
Fig. \ref{fig:3a} and Fig. \ref{fig:2a}, it can be noted that the
imperfect SIC results in degrading the ergodic rates of the DL cell
center user and the UL cell edge user, as their performance rely on
the SIC detection scheme. In addition, from Figs. \ref{fig:3b} and
Fig. \ref{fig:2a} we can see that, the performance of the UL users
degrade greatly as the variance of the residual self interference
rises. 

\begin{figure}[H]
\noindent \begin{centering}
\subfloat[\label{fig:4a}Ergodic rates versus number of STAR-RIS elements $N$
, case I.]{\noindent \begin{centering}
\includegraphics[scale=0.5]{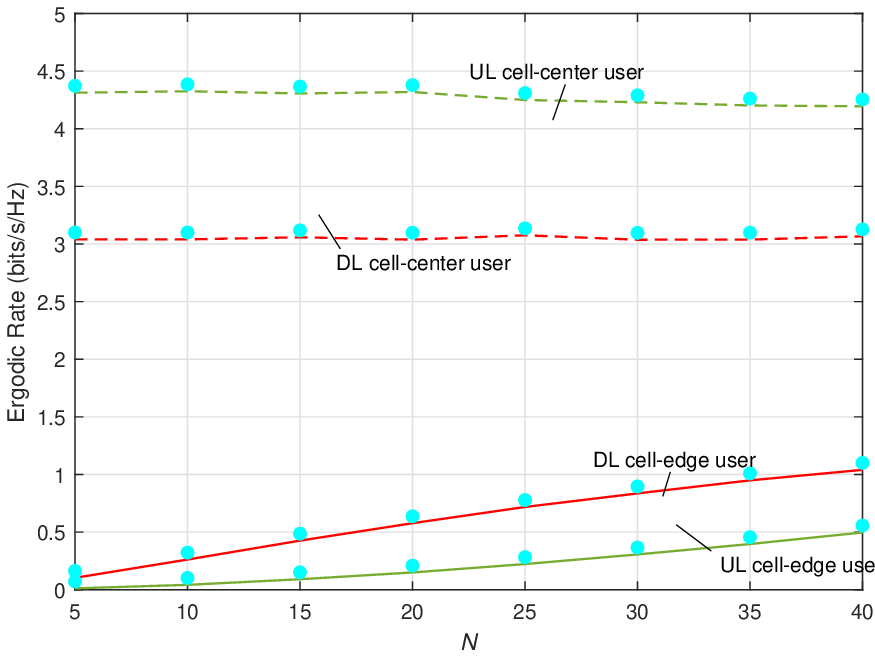}
\par\end{centering}

}
\par\end{centering}

\noindent \begin{centering}
\subfloat[\label{fig:4b}Ergodic rates versus number of STAR-RIS elements $N$,
case II. ]{\noindent \begin{centering}
\includegraphics[scale=0.5]{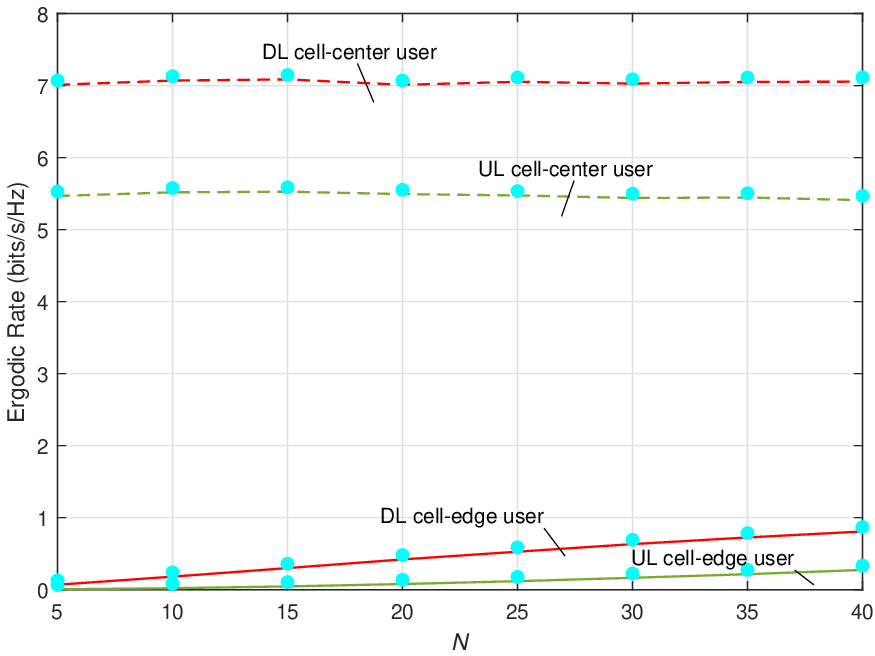}
\par\end{centering}

}
\par\end{centering}

\noindent \begin{centering}
\subfloat[\label{fig:4c}Ergodic sum rates versus number of STAR-RIS elements
$N$. ]{\noindent \begin{centering}
\includegraphics[scale=0.5]{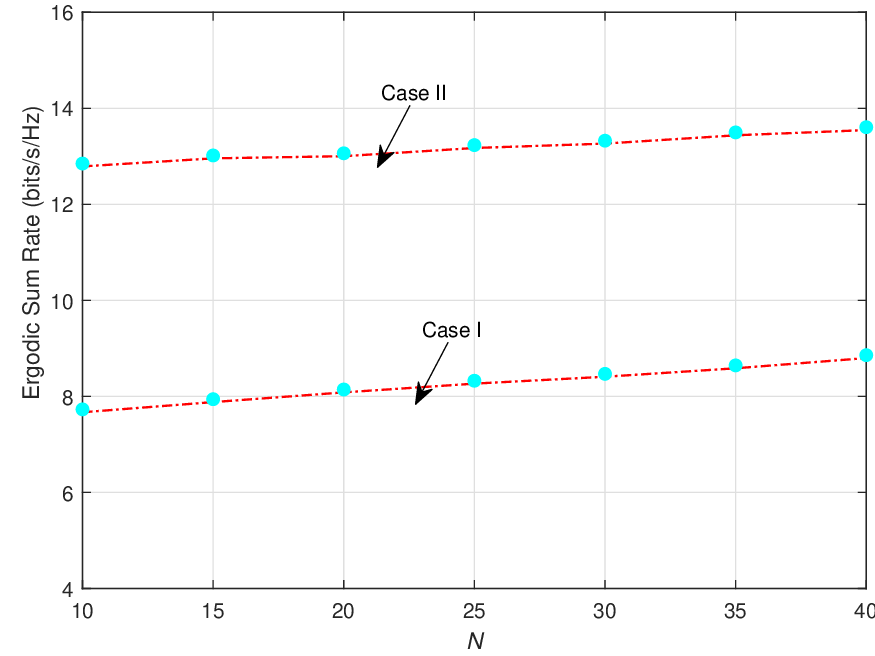}
\par\end{centering}

}
\par\end{centering}

\protect\caption{\label{fig:4}Ergodic rates versus number of STAR-RIS elements $N$
for different system design.}

\end{figure}

Furthermore, in Fig. \ref{fig:4}, we depict the ergodic rates and
sum-rates versus number of STAR-RIS elements, $N$, for different
edge users' target rates, when $\bar{\gamma}=40\textrm{ dB}$. Fig
\ref{fig:4a} represents the high data rate requirement (case I),
and Fig \ref{fig:4b} represents the low data rate requirement (case
II). Firstly, in general, increasing number of the STAR-RIS elements
$N$ enhances the ergodic rates of the edge users, while the performance
of the cell center users is dominant by the direct links between the
BS and the users, and their achievable rates almost fixed with $N$.
Having a closer look at these results, one can observe that, when
the data rate requirements of the edge users are relatively high as
in case I, more power will be allocated to the edge users and less
power to the cell-center users and as a result the total sum rate
will be relatively low as shown in Fig. \ref{fig:4c}. On the other
hand, when the data rate requirements of the edge users are relatively
low as in case II, more power will be allocated to the cell-center
users and this leads to increase the total sum rates as explained
in Fig. \ref{fig:4c}. 

\begin{figure}[H]
\noindent \begin{centering}
\includegraphics[scale=0.6]{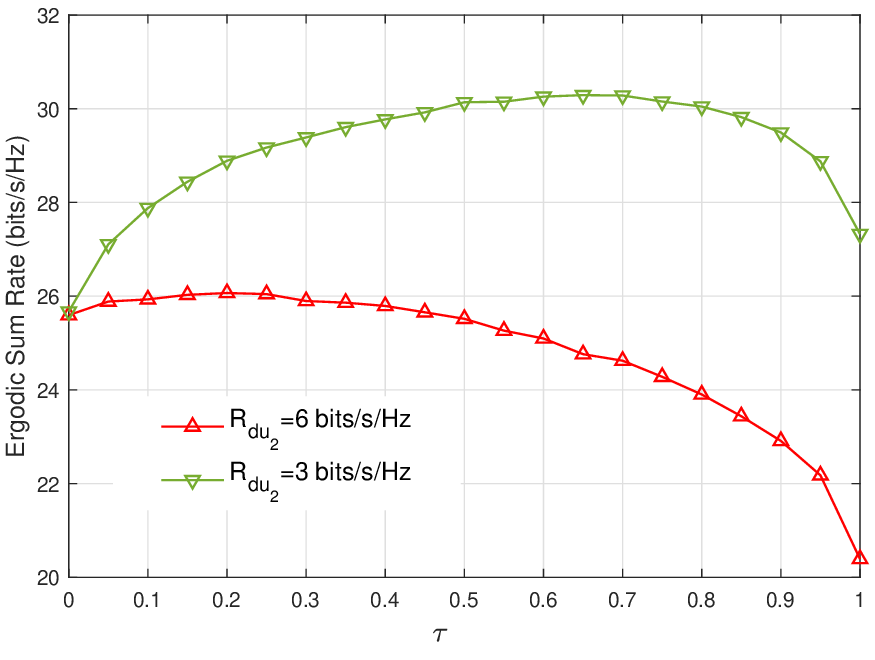}
\par\end{centering}

\protect\caption{\label{fig:5}Ergodic sum-rates versus power allocation coefficient.}
\end{figure}

To explain clearly the impact of the power allocation between the
UL and DL transmissions on the total sum-rate, in Fig. \ref{fig:5}
we plot the ergodic sum-rate versus the power allocation coefficient,
$\tau$, for two different values of the maximum target rate at the
DL edge user 6 bits/s/Hz and 3 bits/s/Hz, when $P_{t}=50\textrm{ dBW}$.
Notably and as expected, when the target rate is high (6 bits/s/Hz)
more power will be allocated to the DL edge user, and less power allocated
to the DL cell center user. As a result, the DL sum rate will be small
and the total sum rate will be dominated by the UL sum rate. For instance,
when $\tau=0$ all power is allocated to the UL transmission and the
achievable sum rate is around 25.6 bits/s/Hz, while when $\tau=1$
the all power is allocated to the DL transmission and the achievable
sum rate is around 20.4 bits/s/Hz, and the optimal $\tau$ is 0.2.
On the other hand, if the target rate at the DL cell edge user is
small (3 bits/s/Hz) more power can be allocated to the DL cell center
user. Thus, the DL sum rate will be high and the total sum rate will
be dominant by both UL and DL. For instance, when $\tau=1$ the achievable
sum rate is now around 27.3 bits/s/Hz, and the optimal $\tau$ in
this case is around 0.65.

Moreover, in Fig. \ref{fig:6} we plot the achievable rates of the
bidirectional communication scenario against number of the STAR-RIS
elements, $N$, for the ideal case ($\Xi=0$, $V=0$), imperfect SIC
($\Xi=0.5$ ) and imperfect SIS ($V=1.25$) schemes, when $\bar{\gamma}=40\textrm{ dB}$.
The ergodic rates of the cell-center user and cell-edge user enhance
with increasing number of the STAR-RIS units. However, the performance
of the cell-center user degrades significantly when the interference
cancellation schemes, SIC and SIS, are imperfect/unideal which is
not the case for the cell-edge user. 

\noindent \begin{center}
\begin{figure}[H]
\noindent \begin{centering}
\includegraphics[scale=0.6]{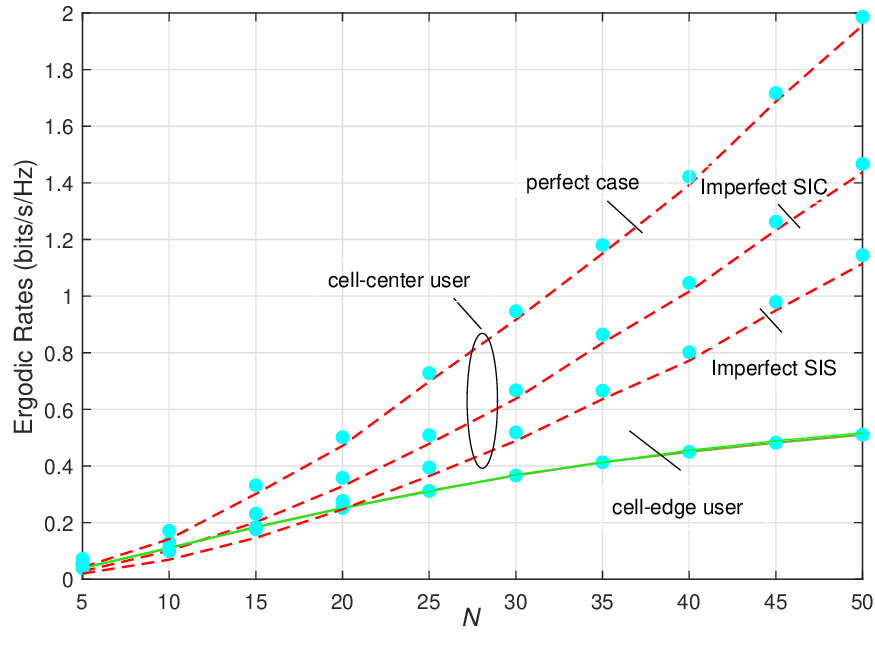}
\par\end{centering}

\protect\caption{\label{fig:6}Ergodic rates of the bidirectional communication versus
number of STAR-RIS elements $N$.}
\end{figure}

\par\end{center}

Fig. \ref{fig:7}, illustrates the ergodic rates of the bidirectional
communication versus the transmit SNR, $\bar{\gamma}$. In these results,
the required data rate of the edge user is higher than that in Fig.
\ref{fig:6}, thus high power has been allocated to the edge user
at the expense of the cell-center user. As expected, from Fig. \ref{fig:7},
the achievable rates of both users can be improved significantly by
increasing the transmit SNR.

\noindent \begin{center}
\begin{figure}[H]
\noindent \begin{centering}
\includegraphics[scale=0.6]{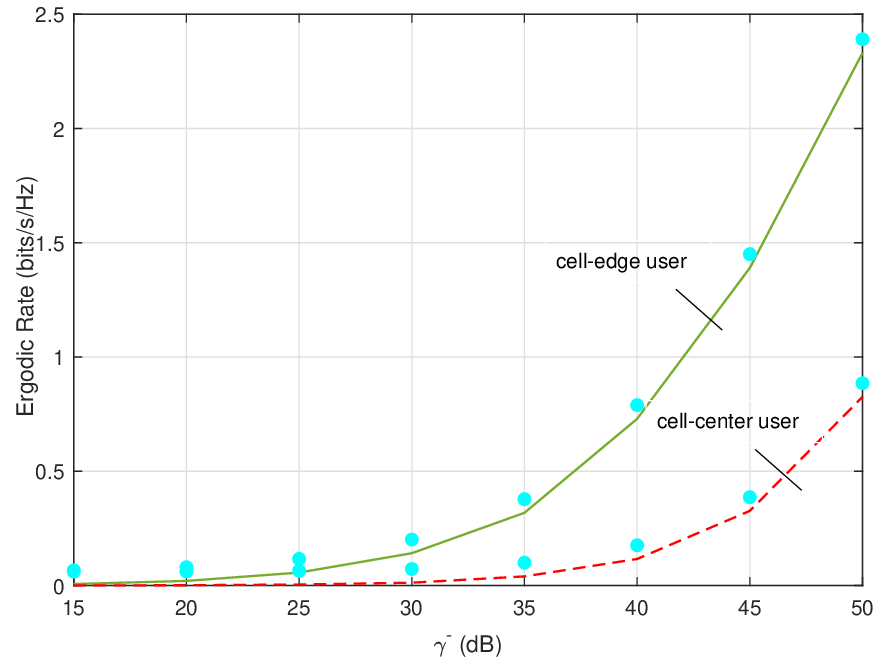}
\par\end{centering}

\protect\caption{\label{fig:7}Ergodic rates of the bidirectional communication versus
transmit SNR $\bar{\gamma}$. }
\end{figure}

\par\end{center}

\begin{figure}[H]
\noindent \begin{centering}
\includegraphics[scale=0.6]{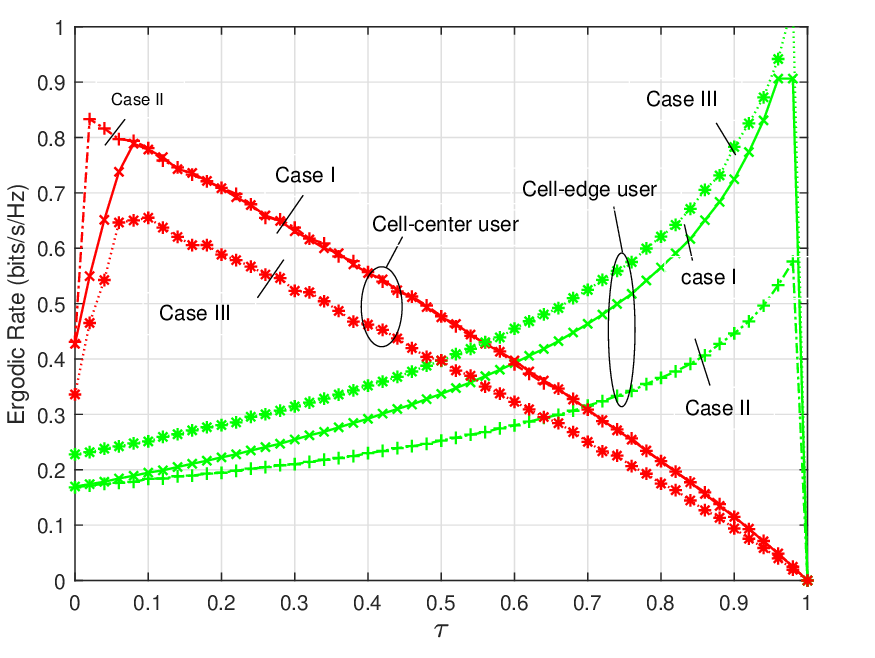}
\par\end{centering}

\protect\caption{\label{fig:8}Ergodic rates of the bidirectional communication versus
power allocation coefficient.}
\end{figure}

Finally, to investigate the impact of the power allocation on the
achievable rates of the bidirectional communication, in Fig. \ref{fig:8}
we plot the ergodic rates versus the power allocation coefficient,
$\tau$, for different cases, case I: $p_{u_{1}}=p_{u_{2}}=0.5P_{u},P_{b_{1}}=0.1P_{b},P_{b_{2}}=0.9P_{b}$,
case II: $p_{u_{1}}=p_{u_{2}}=0.5P_{u},P_{b_{1}}=0.4P_{b},P_{b_{2}}=0.6P_{b}$,
and case III: $p_{u_{1}}=0.6,p_{u_{2}}=0.4P_{u},P_{b_{1}}=0.1P_{b},P_{b_{2}}=0.9P_{b}$,
when $P_{t}=40\textrm{ dBW}$. From these results, we can notice that
the performance of the DL cell-edge user is sensitive to the DL power
where increasing the power allocated to the DL cell-edge user at the
BS enhances the achievable rate at the user. On the other side, the
cell-center user is more sensitive to the UL power and reducing the
power allocated to the UL cell-edge user degrades the data rate. Interestingly
enough, for each user, there exists an optimal power allocation coefficient.
The DL cell-center user reaches the optimal performance with a small
value of $\tau$, e.g., high power allocated to the UL transmission,
while the DL cell-edge user achieves the optimal performance with
a high value of $\tau$, e.g., high power allocated to the DL transmission.

\section{CONCLUSIONS \label{sec:CONCLUSIONs} }

This study investigated a STAR-RIS-assisted FD communication system,
with the STAR-RIS deployed at the cell-edge to support cell-edge users.
We applied a NOMA pairing scheme and derived closed-form expressions
for the ergodic rates. Additionally, our analysis was extended to
encompass bidirectional communication between cell-center and cell-edge
users. Furthermore, we formulated and solved an optimization problem
aimed at maximizing the achievable sum-rate by determining the optimal
STAR-RIS design and power allocation scheme. The results demonstrated
that increasing the transmitted SNR enhances achievable rates, and
using a large number of STAR-RIS elements improves cell-edge user
performance. Additionally, we observed that imperfect SIC degrades
the achievable rates of the DL cell-center user and the UL cell-edge
user, while imperfect SIS significantly impacts UL user performance.

\section*{Appendix A}

By using Jensen inequality, the ergodic rate can be approximated as

\begin{eqnarray*}
\mathcal{E}\left[R_{u_{1d}}\right] & \thickapprox & \log_{2}\left(1+\right.
\end{eqnarray*}

\begin{equation}
\left.\mathcal{E}\left\{ \frac{P_{b_{1}}A_{u_{1d}}}{\Xi P_{b_{2}}A_{u_{1d}}+p_{u_{1u}}C_{u_{1d}}+p_{u_{2u}}D_{u_{1d}}+\sigma_{u_{1d}}^{2}}\right\} \right)
\end{equation}

1- The average of the first term $A_{u_{1d}}$ can be calculated,
after removing the zero expectation terms, as

\[
\mathcal{E}\left\{ \left|\sqrt{l_{b,u_{1d}}^{-m}}h_{b,u_{1d}}+\sqrt{l_{b,r}^{-m}l_{r,u_{1d}}^{-m}}\mathbf{g}_{r,u_{1d}}\Theta\mathbf{g}_{b,r}\right|^{2}\right\} =
\]

\begin{equation}
\mathcal{E}\left\{ l_{b,u_{1d}}^{-m}\left|h_{b,u_{1d}}\right|^{2}\right\} +l_{b,r}^{-m}\mathcal{E}\left\{ l_{r,u_{1d}}^{-m}\right\} \mathcal{E}\left\{ \left|\mathbf{g}_{r,u_{1d}}\Theta\mathbf{g}_{b,r}\right|^{2}\right\} 
\end{equation}

The first expectation, $\mathcal{E}\left\{ l_{b,u_{1d}}^{-m}\left|h_{b,u_{1d}}\right|^{2}\right\} =\mathcal{E}\left\{ l_{b,u_{1d}}^{-m}\right\} $.
The probability density function (PDF) of the strong user at distance
$r$ relative to the BS is $f_{d}\left(r\right)=\frac{2r}{R^{2}}$
\cite{salim11}. Thus we can get,

\[
\mathcal{E}\left\{ l_{b,u_{1d}}^{-m}\right\} =\stackrel[0]{R}{\int}\left(1+r_{b,u_{1d}}\right)^{-m}\frac{2\left(r_{b,u_{1d}}\right)}{\left(R\right)^{2}}dr_{b,u_{1d}}=
\]

\begin{equation}
\frac{2\left(1+R\right)^{-m}\left(-1+R^{2}+mR\left(1+R\right)+\left(1+R\right)^{m}\right)}{\left(m-2\right)\left(m-1\right)R^{2}}\label{eq:42sub}
\end{equation}

\noindent The second expectation, considering the distribution of
the distance between a fixed point outside a circle and a random point
inside the circle in \cite{booknew}, we can write 

\[
\mathcal{E}\left\{ l_{r,u_{1d}}^{-m}\right\} =\stackrel[r_{1}]{r_{1}+2R}{\int}r_{r,u1d}^{-m}\frac{2r_{r,u_{1d}}}{\pi R^{2}}\cos^{-1}\left(\frac{1}{r_{r,u_{1d}}}\right.
\]

\begin{equation}
\left.\left(r_{1}+\frac{\left(r_{r,u_{1d}}^{2}-r_{1}^{2}\right)}{2\left(R+r_{1}\right)}\right)\right)dr_{r,u_{1d}}\label{eq:sup7}
\end{equation}

\noindent where $r_{1}\leq r_{r,u_{1d}}\leq r_{1}+2R$, $r_{1}=d_{b,r}-R$
. Using Gaussian Quadrature rules we can get

\[
\mathcal{E}\left\{ l_{r,u1d}^{-m}\right\} =\stackrel[j=1]{C}{\sum}\textrm{H}_{j}\,\left(1+\left(R\, r_{j}+R\right)\right)^{-m}\frac{2\left(R\, r_{j}+R\right)}{\pi R^{2}}
\]

\begin{equation}
\times\cos^{-1}\left(\frac{1}{\left(R\, r_{j}+R\right)}\left(r_{1}+\frac{\left(\left(R\, r_{j}+R\right)^{2}-r_{1}^{2}\right)}{2\left(R+r_{1}\right)}\right)\right)\label{eq:sup8}
\end{equation}

\noindent where $r_{j}$ and $\textrm{H}_{j}$ are the $j^{th}$ zero
and the weighting factor of the Laguerre polynomials, respectively
\cite{book2}. The last expectation can be written after removing
the zero expectation terms as

\[
\mathscr{E}\left\{ \left|\mathbf{g}_{_{r,u1d}}\Theta\mathbf{g}_{b,r}\right|^{2}\right\} =\frac{\kappa_{r,u1d}}{\kappa_{r,u1d}+1}\frac{\kappa_{b,r}}{\kappa_{b,r}+1}\mathscr{E}\left|\mathbf{\bar{g}}_{_{r,u1d}}\Theta\bar{\mathbf{g}}_{b,r}\right|^{2}+\frac{\kappa_{r,u1d}}{\kappa_{r,u1d}+1}\frac{1}{\kappa_{b,r}+1}\mathscr{E}\left|\mathbf{\bar{g}}_{_{r,u1d}}\Theta\tilde{\mathbf{g}}_{b,r}\right|^{2}
\]

\begin{equation}
+\frac{\kappa_{b,r}}{\kappa_{b,r}+1}\frac{1}{\kappa_{r,u1d}+1}\mathscr{E}\left|\mathbf{\tilde{g}}_{_{r,u1d}}\Theta\bar{\mathbf{g}}_{b,r}\right|^{2}+\frac{1}{\kappa_{r,u1d}+1}\frac{1}{\kappa_{b,r}+1}\mathscr{E}\left|\mathbf{\tilde{g}}_{_{r,u1d}}\Theta\mathbf{\tilde{g}}_{b,r}\right|^{2}\label{eq:sup10}
\end{equation}

Now, the first term in (\ref{eq:sup10}) is 

\[
\mathscr{E}\left|\mathbf{\bar{g}}_{_{r,u1d}}\Theta\bar{\mathbf{g}}_{b,r}\right|^{2}=
\]

\begin{equation}
\left|\stackrel[n=1]{N}{\sum}a_{N,n}\left(\psi_{b,r}^{a},\psi_{b,r}^{e}\right)\rho_{n}^{k}e^{j\phi_{n}^{k}}a_{N,n}\left(\psi_{r,u_{1d}}^{a},\psi_{r,u_{1d}}^{e}\right)\right|^{2}=\xi_{1}
\end{equation}

Similarly, the second term,

\[
\mathscr{E}\left|\mathbf{\bar{g}}_{_{r,u1d}}\Theta\tilde{\mathbf{g}}_{b,r}\right|^{2}=\stackrel[n=1]{N}{\sum}\left|\rho_{n}^{k}\right|^{2}+
\]

\[
\mathscr{E}\left\{ \stackrel[n_{1}=1]{N}{\sum}\stackrel[n_{2}\neq n_{1}]{N}{\sum}\left(a_{Nn_{1}}\left(\psi_{r,u_{1d}}^{a},\psi_{r,u_{1d}}^{e}\right)\rho_{n_{1}}^{k}e^{j\phi_{n_{1}}^{k}}\left[\tilde{\mathbf{g}}_{b,r}\right]_{n_{1}}\right)\right.
\]

\begin{equation}
\left.\left(a_{Nn_{2}}\left(\psi_{r,u_{1d}}^{a},\psi_{r,u_{1d}}^{e}\right)\rho_{n_{2}}^{k}e^{j\phi_{n_{2}}^{k}}\left[\tilde{\mathbf{g}}_{b,r}\right]_{n_{2}}\right)^{H}\right\} =\stackrel[n=1]{N}{\sum}\left|\rho_{n}^{k}\right|^{2}
\end{equation}

\noindent The third term,

\[
\mathscr{E}\left|\mathbf{\tilde{g}}_{_{r,u1d}}\Theta\bar{\mathbf{g}}_{b,r}\right|^{2}=\stackrel[n=1]{N}{\sum}\left|\rho_{n}^{k}\right|^{2}+
\]

\[
\mathscr{E}\left\{ \stackrel[n_{1}=1]{N}{\sum}\stackrel[n_{2}\neq n_{1}]{N}{\sum}\left(\left[\mathbf{\tilde{g}}_{_{r,u1d}}\right]_{n_{1}}\rho_{n_{1}}^{k}e^{j\phi_{n_{1}}^{k}}a_{Nn1}\left(\psi_{b,r}^{a},\psi_{b,r}^{e}\right)\right)\right.
\]

\begin{equation}
\left.\left(\left[\mathbf{\tilde{g}}_{_{r,u1d}}\right]_{n_{2}}\rho_{n_{1}}^{k}e^{j\phi_{n_{1}}^{k}}a_{Nn_{2}}\left(\psi_{b,r}^{a},\psi_{b,r}^{e}\right)\right)^{H}\right\} =\stackrel[n=1]{N}{\sum}\left|\rho_{n}^{k}\right|^{2}
\end{equation}

\noindent The last term,

\[
\mathscr{E}\left|\mathbf{\tilde{g}}_{_{r,u1d}}\Theta\mathbf{\tilde{g}}_{b,r}\right|^{2}=
\]

\begin{equation}
\mathscr{E}\left|\stackrel[n=1]{N}{\sum}\left[\mathbf{\tilde{g}}_{_{r,u1d}}\right]_{n}\rho_{n}^{k}e^{j\phi_{n}^{k}}\left[\mathbf{\tilde{g}}_{b,r}\right]_{n}\right|^{2}=\stackrel[n=1]{N}{\sum}\left|\rho_{n}^{k}\right|^{2}
\end{equation}

Now, we are ready to write the average as 

\[
\mathscr{E}\left\{ \left|\mathbf{g}_{r,u_{1d}}\Theta\mathbf{g}_{b,r}\right|^{2}\right\} =\frac{\kappa_{r,u_{1d}}}{\kappa_{r,u_{1d}}+1}\frac{\kappa_{b,r}}{\kappa_{b,r}+1}\xi_{1}
\]

\[
+\frac{\kappa_{r,u_{1d}}}{\kappa_{r,u_{1d}}+1}\frac{\stackrel[n=1]{N}{\sum}\left|\rho_{n}^{k}\right|^{2}}{\kappa_{b,r}+1}+\frac{\kappa_{b,r}}{\kappa_{b,r}+1}\frac{\stackrel[n=1]{N}{\sum}\left|\rho_{n}^{k}\right|^{2}}{\kappa_{r,u_{1d}}+1}
\]

\begin{equation}
+\frac{1}{\kappa_{r,u_{1d}}+1}\frac{1}{\kappa_{b,r}+1}\stackrel[n=1]{N}{\sum}\left|\rho_{n}^{k}\right|^{2}\label{eq:sub52}
\end{equation}

2- The average of the term $C_{u_{1d}}$ can be calculated by

\[
\mathcal{E}\left\{ \left|\sqrt{l_{u_{1d},u_{1u}}^{-m}}h_{u_{1d},u_{1u}}+\sqrt{l_{r,u_{1u}}^{-m}l_{r,u_{1d}}^{-m}}\mathbf{g}_{r,u_{1d}}\Theta\mathbf{g}_{r,u_{1u}}\right|^{2}\right\} 
\]

\begin{equation}
=\mathcal{E}\left\{ l_{u_{1d},u_{1u}}^{-m}\left|h_{u_{1d},u_{1u}}\right|^{2}\right\} +\mathcal{E}\left\{ l_{r,u_{1u}}^{-m}l_{r,u_{1d}}^{-m}\left|\mathbf{g}_{r,u_{1d}}\Theta\mathbf{g}_{r,u_{1u}}\right|^{2}\right\} 
\end{equation}

\noindent The fist expectation $\mathcal{E}\left\{ l_{u_{1d},u_{1u}}^{-m}\left|h_{u_{1d},u_{1u}}\right|^{2}\right\} =\mathcal{E}\left\{ l_{u_{1d},u_{1u}}^{-m}\right\} $.
Considering the distribution of the distance between two random points
inside a circle in \cite{booknew}, we can write 

\[
\mathcal{E}\left\{ l_{u_{1d},u_{1u}}^{-m}\right\} =\stackrel[r_{0}]{2R}{\int}\left(1+r_{u_{1d},u_{1u}}\right)^{-m}\frac{4r_{u_{1d},u_{1u}}}{\pi R^{2}}\times
\]

\begin{equation}
\left(\cos^{-1}\left(\frac{r_{u_{1d},u_{1u}}}{2R}\right)-\frac{r_{u_{1d},u_{1u}}}{2R}\left(\sqrt{1-\frac{r_{u_{1d},u_{1u}}^{2}}{4R^{2}}}\right)\right)dr_{u_{1d},u_{1u}}
\end{equation}

which can be found as 

\[
\mathcal{E}\left\{ l_{u_{1d},u_{1u}}^{-m}\right\} =\frac{2}{\left(2-3m+m^{2}\right)R^{2}}-
\]

\[
\frac{2\textrm{F}\left(\left\{ \frac{1}{2},-1+\frac{m}{2},-\frac{1}{2}+\frac{m}{2}\right\} ,\left\{ \frac{-1}{2},1\right\} ,4R^{2}\right)}{\left(2-3m+m^{2}\right)R^{2}}
\]

\[
-\textrm{F}\left(\left\{ \frac{3}{2},\frac{1}{2}+\frac{m}{2},\frac{m}{2}\right\} ,\left\{ \frac{1}{2},3\right\} ,4R^{2}\right)
\]

\[
+\frac{64mR\,\textrm{F}\left(\left\{ 2,\frac{1}{2}+\frac{m}{2},1+\frac{m}{2}\right\} ,\left\{ \frac{3}{2},\frac{7}{2}\right\} ,4R^{2}\right)}{15\pi}
\]

\begin{equation}
-\frac{64mR\,\textrm{F}\left(\left\{ 2,\frac{1}{2}+\frac{m}{2},1+\frac{m}{2}\right\} ,\left\{ \frac{5}{2},\frac{5}{2}\right\} ,4R^{2}\right)}{9\pi}
\end{equation}

\noindent where $\textrm{F}\left(.\right)$ is Hypergeometric function.
The second expectation, $\mathcal{E}\left\{ l_{r,u_{1d}}^{-m}\right\} $
is derived in (\ref{eq:sup7}) and (\ref{eq:sup8}). Similar to the
derivation in (\ref{eq:sup7}) and (\ref{eq:sup10}) we can get

\[
\mathcal{E}\left\{ l_{r,u_{1u}}^{-m}\right\} =\stackrel[j=1]{C}{\sum}\textrm{H}_{j}\,\left(1+\left(R\, r_{j}+R\right)\right)^{-m}\frac{2\left(R\, r_{j}+R\right)}{\pi R^{2}}
\]

\begin{equation}
\times\cos^{-1}\left(\frac{1}{\left(R\, r_{j}+R\right)}\left(r_{1}+\frac{\left(\left(R\, r_{j}+R\right)^{2}-r_{1}^{2}\right)}{2\left(R+r_{1}\right)}\right)\right)
\end{equation}

and

\[
\mathscr{E}\left\{ \left|\mathbf{g}_{_{r,u1d}}\Theta\mathbf{g}_{_{r,u1u}}\right|^{2}\right\} =\frac{\kappa_{r,u1d}}{\kappa_{r,u1d}+1}\frac{\kappa_{r,u1u}}{\kappa_{r,u1u}+1}\xi_{2}
\]

\[
+\frac{\kappa_{r,u1d}}{\kappa_{r,u1d}+1}\frac{\stackrel[n=1]{N}{\sum}\left|\rho_{n}^{k}\right|^{2}}{\kappa_{r,u1u}+1}+\frac{\kappa_{r,u1u}}{\kappa_{r,u1u}+1}\frac{\stackrel[n=1]{N}{\sum}\left|\rho_{n}^{k}\right|^{2}}{\kappa_{r,u1d}+1}
\]

\begin{equation}
+\frac{1}{\kappa_{r,u1d}+1}\frac{\stackrel[n=1]{N}{\sum}\left|\rho_{n}^{k}\right|^{2}}{\kappa_{r,u1u}+1}
\end{equation}

\noindent where $\xi_{2}=\left|\bar{\mathbf{g}}_{_{r,u1d}}\Theta\mathbf{\bar{g}}_{_{r,u1u}}\right|^{2}.$

3- The average of the term $D_{u_{1d}}$ can be calculated by

\[
\mathcal{E}\left\{ l_{r,u_{2u}}^{-m}l_{r,u_{1d}}^{-m}\left|\mathbf{g}_{r,u_{1d}}\Theta\mathbf{g}_{r,u_{2u}}\right|^{2}\right\} =
\]

\begin{equation}
\mathcal{E}\left\{ l_{r,u_{2u}}^{-m}\right\} \mathcal{E}\left\{ l_{r,u_{1d}}^{-m}\right\} \mathcal{E}\left\{ \left|\mathbf{g}_{r,u_{1d}}\Theta\mathbf{g}_{r,u_{2u}}\right|^{2}\right\} 
\end{equation}

The PDF of the weak user at radius $r_{r}$ relative to the RIS is
$f_{d}\left(r_{r}\right)=\frac{2r_{r}}{R_{r}^{2}},$ \cite{salim11}.
Thus, we can get

\[
\mathcal{E}\left\{ l_{r,u_{2u}}^{-m}\right\} =\frac{2\left(1+R_{r}\right)^{-m}}{\left(m-2\right)\left(m-1\right)R_{r}^{2}}
\]

\begin{equation}
\times\left(-1+R_{r}^{2}+mR_{r}\left(1+R_{r}\right)+\left(1+R_{r}\right)^{m}\right)
\end{equation}

\noindent where $\mathcal{E}\left\{ l_{r,u_{1d}}^{-m}\right\} $ is
derived in (\ref{eq:sup7}) and (\ref{eq:sup8}). Similar to the derivation
in (\ref{eq:sup10}) we can obtain

\[
\mathscr{E}\left\{ \left|\mathbf{g}_{_{r,u1d}}\Theta\mathbf{g}_{u_{2u,r}}\right|^{2}\right\} =\frac{\kappa_{r,u1d}}{\kappa_{r,u1d}+1}\frac{\kappa_{r,u_{2u}}}{\kappa_{r,u_{2u}}+1}\xi_{3}+
\]

\[
\frac{\kappa_{r,u1d}}{\kappa_{r,u1d}+1}\frac{\stackrel[n=1]{N}{\sum}\left|\rho_{n}^{k}\right|^{2}}{\kappa_{r,u_{2u}}+1}+\frac{\kappa_{r,u_{2u}}}{\kappa_{r,u_{2u}}+1}\frac{\stackrel[n=1]{N}{\sum}\left|\rho_{n}^{k}\right|^{2}}{\kappa_{r,u1d}+1}
\]

\begin{equation}
+\frac{1}{\kappa_{r,u1d}+1}\frac{\stackrel[n=1]{N}{\sum}\left|\rho_{n}^{k}\right|^{2}}{\kappa_{r,u_{2u}}+1}
\end{equation}

\noindent where $\xi_{3}=\left|\bar{\mathbf{g}}_{_{r,u1d}}\Theta\mathbf{\bar{g}}_{_{r,u2u}}\right|^{2}.$

\section*{Appendix B}

By using Jensen inequality, the ergodic rate can be expressed as

\begin{equation}
\mathcal{E}\left[R_{u_{2d}}\right]\approx\log_{2}\left(1+\mathcal{E}\left[\frac{P_{b_{2}}A_{u_{2d}}}{P_{b_{1}}A_{u_{2d}}+p_{u_{1u}}C_{u_{2d}}+p_{u_{2u}}D_{u_{2d}}+\sigma_{u_{2d}}^{2}}\right]\right)
\end{equation}

1- The average of the first term $A_{u_{2d}}$ can be calculated by
\[
\mathcal{E}\left\{ l_{b,r}^{-m}l_{r,u2d}^{-m}\left|\mathbf{g}_{_{r,u2d}}\Theta_{r}\mathbf{g}_{b,r}\right|^{2}\right\} =
\]

\begin{equation}
l_{b,r}^{-m}\mathcal{E}\left\{ l_{r,u2d}^{-m}\right\} \mathcal{E}\left\{ \left|\mathbf{g}_{_{r,u2d}}\Theta_{r}\mathbf{g}_{b,r}\right|^{2}\right\} 
\end{equation}

\noindent The PDF of the weak user at radius $r_{r}$ relative to
the RIS is $f_{d}\left(r_{r}\right)=\frac{2r_{r}}{R_{r}^{2}},$ \cite{salim11}.
Thus,

\[
\mathcal{E}\left\{ l_{r,u2d}^{-m}\right\} =\stackrel[0]{R_{r}}{\int}\left(1+r_{r,u2d}\right)^{-m}\frac{2\left(r_{r,u2d}\right)}{R_{r}^{2}}dr_{r,u2d}=
\]

\[
\frac{2\left(1+R_{r}\right)^{-m}}{\left(m-2\right)\left(m-1\right)R_{r}^{2}}\times
\]

\begin{equation}
\left(R_{r}^{2}+mR_{r}\left(1+R_{r}\right)+\left(1+R_{r}\right)^{m}-1\right)\label{eq:sup36}
\end{equation}

\noindent Now the average of the other term can be derived as in Appendix
A as,

\[
\mathscr{E}\left\{ \left|\mathbf{g}_{_{r,u2d}}\Theta\mathbf{g}_{b,r}\right|^{2}\right\} =\frac{\kappa_{r,u2d}}{\kappa_{r,u2d}+1}\frac{\kappa_{b,r}}{\kappa_{b,r}+1}\xi_{4}
\]

\[
+\frac{\kappa_{r,u2d}}{\kappa_{r,u2d}+1}\frac{\stackrel[n=1]{N}{\sum}\left|\rho_{n}^{k}\right|^{2}}{\kappa_{b,r}+1}+\frac{\kappa_{b,r}}{\kappa_{b,r}+1}\frac{\stackrel[n=1]{N}{\sum}\left|\rho_{n}^{k}\right|^{2}}{\kappa_{r,u2d}+1}
\]

\begin{equation}
+\frac{1}{\kappa_{r,u2d}+1}\frac{1}{\kappa_{b,r}+1}\stackrel[n=1]{N}{\sum}\left|\rho_{n}^{k}\right|^{2}
\end{equation}

\noindent where $\xi_{4}=\left|\mathbf{\bar{g}}_{_{r,u2d}}\Theta\bar{\mathbf{g}}_{b,r}\right|^{2}.$

2- The average of the second term, $C_{u_{2d}}$, can be calculated
by

\[
\mathcal{E}\left\{ l_{r,u2d}^{-m}l_{r,u1u}^{-m}\left|\mathbf{g}_{_{r,u2d}}\Theta_{k}\mathbf{g}_{r,u_{1u}}\right|^{2}\right\} =
\]

\begin{equation}
\mathcal{E}\left\{ l_{r,u2d}^{-m}\right\} \mathcal{E}\left\{ l_{r,u1u}^{-m}\right\} \mathcal{E}\left\{ \left|\mathbf{g}_{_{r,u2d}}\Theta_{k}\mathbf{g}_{r,u_{1u}}\right|^{2}\right\} 
\end{equation}

\noindent The term $\mathcal{E}\left\{ l_{r,u2d}^{-m}\right\} $ is
derived in (\ref{eq:sup36}). Considering the distribution of the
distance between a fixed point outside and a random point inside a
circle in \cite{booknew}, we can obtain the aveage of the second
term as 

\[
\mathcal{E}\left\{ l_{r,u1u}^{-m}\right\} =\stackrel[r_{1}]{r_{1}+2R}{\int}r_{r,u1u}^{-m}\frac{2r_{r,u1u}}{\pi R^{2}}\times
\]

\begin{equation}
\cos^{-1}\left(\frac{1}{r_{r,u1u}}\left(r_{1}+\frac{\left(r_{r,u1u}^{2}-r_{1}^{2}\right)}{2\left(R+r_{1}\right)}\right)\right)dr_{r,u1u}\label{eq:sup39}
\end{equation}

\noindent where $r_{1}\leq r_{r,u1u}\leq r_{1}+2R$, $r_{1}=d_{b,r}-R$
is the distance from RIS to the boundary. The last expression can
be rewritten as,

\[
\mathcal{E}\left\{ l_{r,u1u}^{-m}\right\} =\stackrel[j=1]{C}{\sum}\textrm{H}_{j}\,\left(1+\left(R\, r_{j}+R\right)\right)^{-m}\frac{2\left(R\, r_{j}+R\right)}{\pi R^{2}}
\]

\begin{equation}
\cos^{-1}\left(\frac{1}{\left(R\, r_{j}+R\right)}\left(r_{1}+\frac{\left(\left(R\, r_{j}+R\right)^{2}-r_{1}^{2}\right)}{2\left(R+r_{1}\right)}\right)\right)\label{eq:sup40}
\end{equation}

The average of the last term is

\[
\mathscr{E}\left\{ \left|\mathbf{g}_{_{r,u2d}}\Theta_{k}\mathbf{g}_{r,u_{1u}}\right|^{2}\right\} =\frac{\kappa_{r,u2d}}{\kappa_{r,u2d}+1}\frac{\kappa_{r,u_{1u}}}{\kappa_{r,u_{1u}}+1}\xi_{5}+
\]

\[
+\frac{\kappa_{r,u2d}}{\kappa_{r,u2d}+1}\frac{\stackrel[n=1]{N}{\sum}\left|\rho_{n}^{k}\right|^{2}}{\kappa_{r,u_{1u}}+1}+\frac{\kappa_{u_{1u,r}}}{\kappa_{u_{1u,r}}+1}\frac{\stackrel[n=1]{N}{\sum}\left|\rho_{n}^{k}\right|^{2}}{\kappa_{r,u2d}+1}
\]

\begin{equation}
+\frac{1}{\kappa_{r,u2d}+1}\frac{1}{\kappa_{r,u_{1u}}+1}\stackrel[n=1]{N}{\sum}\left|\rho_{n}^{k}\right|^{2}
\end{equation}

\noindent where $\xi_{5}=\left|\mathbf{\bar{g}}_{_{r,u2d}}\Theta\bar{\mathbf{g}}_{r,u_{1u}}\right|^{2}.$

3- The average of the third term, $D_{u_{2d}}$, can be derived as 

\[
\mathcal{E}\left\{ l_{r,u2d}^{-m}l_{r,u2u}^{-m}\left|\mathbf{g}_{_{r,u2d}}\Theta\mathbf{g}_{r,u_{2u}}\right|^{2}\right\} =
\]

\begin{equation}
\mathcal{E}\left\{ l_{r,u2d}^{-m}\right\} \mathcal{E}\left\{ l_{r,u2u}^{-m}\right\} \mathcal{E}\left\{ \left|\mathbf{g}_{_{r,u2d}}\Theta\mathbf{g}_{r,u_{2u}}\right|^{2}\right\} 
\end{equation}

\noindent The term $\mathcal{E}\left\{ l_{r,u2d}^{-m}\right\} $ is
derived in (\ref{eq:sup36}). Following the derivation in (\ref{eq:sup36})
we can get

\[
\mathcal{E}\left\{ l_{r,u_{2u}}^{-m}\right\} =\frac{2\left(1+R_{r}\right)^{-m}}{\left(m-2\right)\left(m-1\right)R_{r}^{2}}\times
\]

\begin{equation}
\left(-1+R_{r}^{2}+mR_{r}\left(1+R_{r}\right)+\left(1+R_{r}\right)^{m}\right)\label{eq:53}
\end{equation}

and

\[
\mathscr{E}\left\{ \left|\mathbf{g}_{_{r,u2d}}\Theta\mathbf{g}_{r,u_{2u}}\right|^{2}\right\} =\frac{\kappa_{r,u2d}}{\kappa_{r,u2d}+1}\frac{\kappa_{r,u_{2u}}}{\kappa_{r,u_{2u}}+1}\xi_{6}+
\]

\[
\frac{\kappa_{r,u2d}}{\kappa_{r,u2d}+1}\frac{\stackrel[n=1]{N}{\sum}\left|\rho_{n}^{k}\right|^{2}}{\kappa_{r,u_{2u}}+1}+\frac{\kappa_{r,u_{1u}}}{\kappa_{r,u_{1u}}+1}\frac{\stackrel[n=1]{N}{\sum}\left|\rho_{n}^{k}\right|^{2}}{\kappa_{r,u_{2u}}+1}
\]

\begin{equation}
+\frac{1}{\kappa_{r,u2d}+1}\frac{1}{\kappa_{r,u_{2u}}+1}\stackrel[n=1]{N}{\sum}\left|\rho_{n}^{k}\right|^{2}
\end{equation}

\noindent where $\xi_{6}=\left|\mathbf{\bar{g}}_{_{r,u2d}}\Theta\bar{\mathbf{g}}_{r,u_{2u}}\right|^{2}.$

\section*{Appendix C}

By using Jensen inequality, the ergodic rate can be expressed as

\begin{equation}
\mathcal{E}\left[R_{u_{1u}}\right]\approx\log_{2}\left(1+\mathcal{E}\left\{ \frac{p_{u_{1u}}A_{u_{1u}}}{p_{u_{2u}}B_{u_{1u}}+P_{b}C_{u_{1u}}+\left|\tilde{s}\right|^{2}+\sigma_{b}^{2}}\right\} \right)
\end{equation}

1- The average of the first term, $A_{u_{1u}}$, can be derived as

\[
\mathcal{E}\left\{ \left|\sqrt{l_{b,u_{1u}}^{-m}}h_{b,u_{1u}}+\sqrt{l_{b,r}^{-m}l_{r,u1u}^{-m}}\mathbf{g}_{b,r}\Theta\mathbf{g}_{_{r,u1u}}\right|^{2}\right\} =
\]

\begin{equation}
p_{u_{1u}}\mathcal{E}\left|\sqrt{l_{b,u_{1u}}^{-m}}h_{b,u_{1u}}\right|^{2}+\mathcal{E}\left|\sqrt{l_{b,r}^{-m}l_{r,u1u}^{-m}}\mathbf{g}_{b,r}\Theta\mathbf{g}_{_{r,u1u}}\right|^{2}
\end{equation}

The first expectation, $\mathcal{E}\left\{ l_{b,u_{1u}}^{-m}\left|h_{b,u_{1u}}\right|^{2}\right\} =\mathcal{E}\left\{ l_{b,u_{1u}}^{-m}\right\} $.
The PDF of the weak user at radius $r$ relative to the RIS is $f_{d}\left(r\right)=\frac{2r}{R^{2}}$,
\cite{salim11}. Thus,

\[
\mathcal{E}\left\{ l_{b,u_{1u}}^{-m}\right\} =\stackrel[0]{R}{\int}\left(1+r_{b,u_{1u}}\right)^{-m}\frac{2\left(r_{b,u_{1u}}\right)}{\left(R\right)^{2}}dr_{b,u_{1u}}
\]

\[
=\frac{2\left(1+R\right)^{-m}}{\left(m-2\right)\left(m-1\right)R^{2}}\times
\]

\begin{equation}
\left(R^{2}+mR\left(1+R\right)+\left(1+R\right)^{m}-1\right)\label{eq:57}
\end{equation}

\noindent The second term,

\[
\mathcal{E}\left\{ \left|\sqrt{l_{b,r}^{-m}l_{r,u1u}^{-m}}\mathbf{g}_{b,r}\Theta\mathbf{g}_{_{r,u1u}}\right|^{2}\right\} =
\]

\begin{equation}
l_{b,r}^{-m}\mathcal{E}\left\{ l_{r,u1u}^{-m}\left|\mathbf{g}_{b,r}\Theta\mathbf{g}_{_{r,u1u}}\right|^{2}\right\} 
\end{equation}

\noindent in which $\mathcal{E}\left\{ l_{r,u1u}^{-m}\right\} $ is
derived in (\ref{eq:sup39}) and (\ref{eq:sup40}). The average of
the last term can be found as,

\[
\mathscr{E}\left\{ \left|\mathbf{g}_{b,r}\Theta\mathbf{g}_{_{r,u1u}}\right|^{2}\right\} =\frac{\kappa_{r,u1u}}{\kappa_{r,u1u}+1}\frac{\kappa_{b,r}}{\kappa_{b,r}+1}\xi_{7}+
\]

\[
\frac{\kappa_{r,u1u}}{\kappa_{r,u1u}+1}\frac{\stackrel[n=1]{N}{\sum}\left|\rho_{n}^{k}\right|^{2}}{\kappa_{b,r}+1}+\frac{\kappa_{b,r}}{\kappa_{b,r}+1}\frac{\stackrel[n=1]{N}{\sum}\left|\rho_{n}^{k}\right|^{2}}{\kappa_{r,u1u}+1}
\]

\begin{equation}
+\frac{1}{\kappa_{r,u1u}+1}\frac{1}{\kappa_{b,r}+1}\stackrel[n=1]{N}{\sum}\left|\rho_{n}^{k}\right|^{2}
\end{equation}

\noindent where $\xi_{7}=\left|\bar{\mathbf{g}}_{b,r}\Theta\mathbf{\bar{g}}_{_{r,u1u}}\right|^{2}.$

2- The average of the second term, $B_{u_{1u}}$, can be obtained
as

\[
\mathcal{E}\left\{ l_{b,r}^{-m}l_{r,u2u}^{-m}\left|\mathbf{g}_{b,r}\Theta\mathbf{g}_{_{r,u2u}}\right|^{2}\right\} =
\]

\begin{equation}
l_{b,r}^{-m}\mathcal{E}\left\{ l_{r,u2u}^{-m}\right\} \mathcal{E}\left\{ \left|\mathbf{g}_{b,r}\Theta\mathbf{g}_{_{r,u2u}}\right|^{2}\right\} 
\end{equation}

\noindent in which $\mathcal{E}\left\{ l_{r,u2u}^{-m}\right\} $ is
derived in (\ref{eq:53}). The average of the other term can be calculated
as

\[
\mathscr{E}\left\{ \left|\mathbf{g}_{b,r}\Theta\mathbf{g}_{_{r,u2u}}\right|^{2}\right\} =\frac{\kappa_{r,u2u}}{\kappa_{r,u2u}+1}\frac{\kappa_{b,r}}{\kappa_{b,r}+1}\xi_{8}+
\]

\[
\frac{\kappa_{r,u2u}}{\kappa_{r,u2u}+1}\frac{\stackrel[n=1]{N}{\sum}\left|\rho_{n}^{k}\right|^{2}}{\kappa_{b,r}+1}+\frac{\kappa_{b,r}}{\kappa_{b,r}+1}\frac{\stackrel[n=1]{N}{\sum}\left|\rho_{n}^{k}\right|^{2}}{\kappa_{r,u2u}+1}
\]

\begin{equation}
+\frac{1}{\kappa_{r,u2u}+1}\frac{1}{\kappa_{b,r}+1}\stackrel[n=1]{N}{\sum}\left|\rho_{n}^{k}\right|^{2}
\end{equation}

\noindent where $\xi_{8}=\left|\bar{\mathbf{g}}_{b,r}\Theta\mathbf{\bar{g}}_{_{_{r,u2u}}}\right|^{2}.$

3- The average of the third term, $C_{u_{1u}}$ can be derived as

\[
\mathcal{E}\left\{ l_{b,r}^{-m}l_{b,r}^{-m}\left|\mathbf{g}_{b,r}\Theta_{k}\mathbf{g}_{b,r}^{H}\right|^{2}\right\} =
\]

\begin{equation}
l_{b,r}^{-m}l_{b,r}^{-m}\mathcal{E}\left\{ \left|\mathbf{g}_{b,r}\Theta_{k}\mathbf{g}_{b,r}^{H}\right|^{2}\right\} 
\end{equation}

\noindent The average can be written as 

\[
\mathscr{E}\left\{ \left|\mathbf{g}_{b,r}\Theta_{k}\mathbf{g}_{b,r}^{H}\right|^{2}\right\} =\mathscr{E}\left\{ \left|\left(\sqrt{\frac{\kappa_{b,r}}{\kappa_{b,r}+1}}\sqrt{\frac{\kappa_{b,r}}{\kappa_{b,r}+1}}\mathbf{\bar{g}}_{b,r}\Theta\bar{\mathbf{g}}_{b,r}^{H}+\sqrt{\frac{\kappa_{b,r}}{\kappa_{b,r}+1}}\sqrt{\frac{1}{\kappa_{b,r}+1}}\mathbf{\bar{g}}_{b,r}\Theta\mathbf{\tilde{g}}_{b,r}^{H}\right.\right.\right.
\]

\begin{equation}
\left.\left.\left.+\sqrt{\frac{\kappa_{b,r}}{\kappa_{b,r}+1}}\sqrt{\frac{1}{\kappa_{b,r}+1}}\mathbf{\tilde{g}}_{b,r}\Theta\bar{\mathbf{g}}_{b,r}^{H}+\sqrt{\frac{1}{\kappa_{b,r}+1}}\sqrt{\frac{1}{\kappa_{b,r}+1}}\mathbf{\tilde{g}}_{b,r}\Theta\mathbf{\tilde{g}}_{b,r}^{H}\right)\right|^{2}\right\} \label{eq:79}
\end{equation}

\[
\mathscr{E}\left\{ \left|\mathbf{g}_{b,r}\Theta_{k}\mathbf{g}_{b,r}^{H}\right|^{2}\right\} =\left(\frac{\kappa_{b,r}}{\kappa_{b,r}+1}\right)^{2}\mathscr{E}\left|\mathbf{\bar{g}}_{b,r}\Theta\bar{\mathbf{g}}_{b,r}^{H}\right|^{2}+\frac{\kappa_{b,r}}{\kappa_{b,r}+1}\frac{1}{\kappa_{b,r}+1}\mathscr{E}\left|\mathbf{\bar{g}}_{b,r}\Theta\mathbf{\tilde{g}}_{b,r}^{H}\right|^{2}
\]

\begin{equation}
+\frac{\kappa_{b,r}}{\kappa_{b,r}+1}\frac{1}{\kappa_{b,r}+1}\mathscr{E}\left|\mathbf{\tilde{g}}_{b,r}\Theta\bar{\mathbf{g}}_{b,r}^{H}\right|^{2}+\left(\frac{1}{\kappa_{b,r}+1}\right)^{2}\mathscr{E}\left|\mathbf{\tilde{g}}_{b,r}\Theta\mathbf{\tilde{g}}_{b,r}^{H}\right|^{2}+2\frac{\kappa_{b,r}}{\kappa_{b,r}+1}\frac{1}{\kappa_{b,r}+1}\mathscr{E}\left(\mathbf{\bar{g}}_{b,r}\Theta\bar{\mathbf{g}}_{b,r}^{H}\mathbf{\tilde{g}}_{b,r}^{H}\Theta^{H}\mathbf{\tilde{g}}_{b,r}\right)\label{eq:80}
\end{equation}

\noindent The first term in (\ref{eq:80}) can be found as $\mathscr{E}\left|\mathbf{\bar{g}}_{b,r}\Theta\bar{\mathbf{g}}_{b,r}^{H}\right|^{2}=\left|\mathbf{\bar{g}}_{b,r}\Theta\bar{\mathbf{g}}_{b,r}^{H}\right|^{2}=\xi_{9}$.
Similarly, the second term,

\[
\mathbf{\bar{g}}_{b,r}\Theta\mathbf{\tilde{g}}_{b,r}^{H}=\mathbf{a}_{N}\left(\psi_{b,r}^{a},\psi_{b,r}^{e}\right)\Theta\tilde{\mathbf{g}}_{b,r}^{H}
\]

\begin{equation}
=\stackrel[n=1]{N}{\sum}a_{Nn}\left(\psi_{b,r}^{a},\psi_{b,r}^{e}\right)\rho_{n}^{k}e^{j\phi_{n}^{k}}\left[\tilde{\mathbf{g}}_{b,r}\right]_{n}
\end{equation}

\[
\mathscr{E}\left|\mathbf{\bar{g}}_{b,r}\Theta\mathbf{\tilde{g}}_{b,r}^{H}\right|^{2}=\stackrel[n=1]{N}{\sum}\left|\rho_{n}^{k}\right|^{2}+
\]

\[
\mathscr{E}\left\{ \stackrel[n_{1}=1]{N}{\sum}\stackrel[n_{2}\neq n_{1}]{N}{\sum}\left(a_{Nn_{1}}\left(\psi_{b,r}^{a},\psi_{b,r}^{e}\right)\rho_{n_{1}}^{k}e^{j\phi_{n_{1}}^{k}}\left[\tilde{\mathbf{g}}_{b,r}\right]_{n_{1}}\right)\right.
\]

\begin{equation}
\left.\times\left(a_{Nn_{2}}\left(\psi_{b,r}^{a},\psi_{b,r}^{e}\right)\rho_{n_{2}}^{k}e^{j\phi_{n_{2}}^{k}}\left[\tilde{\mathbf{g}}_{b,r}\right]_{n_{2}}\right)^{H}\right\} =\stackrel[n=1]{N}{\sum}\left|\rho_{n}^{k}\right|^{2}
\end{equation}

\noindent The other terms, $\mathbf{\tilde{g}}_{b,r}\Theta\bar{\mathbf{g}}_{b,r}^{H}=\stackrel[n=1]{N}{\sum}\left[\mathbf{\tilde{g}}_{b,r}\right]_{n}\rho_{n}^{k}e^{j\phi_{n}^{k}}a_{N,n}\left(\psi_{b,r}^{a},\psi_{b,r}^{e}\right)$,
so we can write,

\[
\mathscr{E}\left|\mathbf{\tilde{g}}_{b,r}\Theta\bar{\mathbf{g}}_{b,r}^{H}\right|^{2}=\stackrel[n=1]{N}{\sum}\left|\rho_{n}^{k}\right|^{2}+
\]

\[
\mathscr{E}\left\{ \stackrel[n_{1}=1]{N}{\sum}\stackrel[n_{2}\neq n_{1}]{N}{\sum}\left(\left[\mathbf{\tilde{g}}_{b,r}\right]_{n_{1}}\rho_{n_{1}}^{k}e^{j\phi_{n_{1}}^{k}}a_{Nn1}\left(\psi_{b,r}^{a},\psi_{b,r}^{e}\right)\right)\right.
\]

\begin{equation}
\left.\times\left(\left[\mathbf{\tilde{g}}_{b,r}\right]_{n_{2}}\rho_{n_{1}}^{k}e^{j\phi_{n_{1}}^{k}}a_{Nn_{2}}\left(\psi_{b,r}^{a},\psi_{b,r}^{e}\right)\right)^{H}\right\} =\stackrel[n=1]{N}{\sum}\left|\rho_{n}^{k}\right|^{2}
\end{equation}

\noindent and $\mathbf{\tilde{g}}_{b,r}\Theta\mathbf{\tilde{g}}_{b,r}^{H}=\stackrel[n=1]{N}{\sum}\left[\mathbf{\tilde{g}}_{b,r}\right]_{n}\rho_{n}^{k}e^{j\phi_{n}^{k}}\left[\mathbf{\tilde{g}}_{b,r}^{H}\right]_{n}$,
so we can get

\begin{equation}
\mathscr{E}\left|\mathbf{\tilde{g}}_{b,r}\Theta\mathbf{\tilde{g}}_{b,r}^{H}\right|^{2}=\mathscr{E}\left|\stackrel[n=1]{N}{\sum}\left[\mathbf{\tilde{g}}_{b,r}\right]_{n}\rho_{n}^{k}e^{j\phi_{n}^{k}}\left[\mathbf{\tilde{g}}_{b,r}^{H}\right]_{n}\right|^{2}
\end{equation}

\[
\mathscr{E}\left|\mathbf{\tilde{g}}_{b,r}\Theta\mathbf{\tilde{g}}_{b,r}^{H}\right|^{2}=2\stackrel[n=1]{N}{\sum}\left|\rho_{n}^{k}\right|^{2}+
\]

\[
\mathscr{E}\left\{ \stackrel[n_{1}=1]{N}{\sum}\stackrel[n_{2}\neq n_{1}]{N}{\sum}\left(\left[\tilde{\mathbf{g}}_{b,r}\right]_{n_{1}}\rho_{n_{1}}^{k}e^{j\phi_{n_{1}}^{k}}\left[\tilde{\mathbf{g}}_{b,r}^{H}\right]_{n_{1}}\right)\right.
\]

\[
\left.\left(\left[\tilde{\mathbf{g}}_{b,r}\right]_{n_{2}}\rho_{n_{2}}^{k}e^{j\phi_{n_{2}}^{k}}\left[\tilde{\mathbf{g}}_{b,r}^{H}\right]_{n_{2}}\right)^{H}\right\} 
\]

\begin{equation}
=2\stackrel[n=1]{N}{\sum}\left|\rho_{n}^{k}\right|^{2}+\stackrel[n_{1}=1]{N}{\sum}\stackrel[n_{2}\neq n_{1}]{N}{\sum}\left(\rho_{n_{1}}^{k}e^{j\phi_{n_{1}}^{k}}\right)\left(\rho_{n_{2}}^{k}e^{j\phi_{n_{2}}^{k}}\right)^{H}
\end{equation}

The last term

\[
\mathbf{\bar{g}}_{b,r}\Theta\bar{\mathbf{g}}_{b,r}^{H}\mathbf{\tilde{g}}_{b,r}\Theta^{H}\mathbf{\tilde{g}}_{b,r}^{H}=
\]

\begin{equation}
\stackrel[n=1]{N}{\sum}a_{Nn}\left(\psi_{b,r}^{a},\psi_{b,r}^{e}\right)^{H}\rho_{n}^{k}e^{j\phi_{n}^{k}}a_{Nn}\left(\psi_{b,r}^{a},\psi_{b,r}^{e}\right)\stackrel[n=1]{N}{\sum}\left(\rho_{n}^{k}e^{j\phi_{n}^{k}}\right)^{H}
\end{equation}

Now we can write the average as,

\[
\mathscr{E}\left\{ \left|\mathbf{g}_{b,r}\Theta_{k}\mathbf{g}_{b,r}^{H}\right|^{2}\right\} =\left(\frac{\kappa_{b,r}}{\kappa_{b,r}+1}\right)^{2}\xi_{9}
\]

\[
+2\frac{\kappa_{b,r}}{\kappa_{b,r}+1}\frac{1}{\kappa_{b,r}+1}\stackrel[n=1]{N}{\sum}\left|\rho_{n}^{k}\right|^{2}+\left(\frac{1}{\kappa_{b,r}+1}\right)^{2}
\]

\[
\times\left(2\stackrel[n=1]{N}{\sum}\left|\rho_{n}^{k}\right|^{2}+\right.
\]

\[
\left.\stackrel[n_{1}=1]{N}{\sum}\stackrel[n_{2}\neq n_{1}]{N}{\sum}\left(\rho_{n_{1}}^{k}e^{j\phi_{n_{1}}^{k}}\right)\left(\rho_{n_{2}}^{k}e^{j\phi_{n_{2}}^{k}}\right)^{H}\right)
\]

\begin{equation}
+2\frac{\kappa_{b,r}}{\kappa_{b,r}+1}\frac{1}{\kappa_{b,r}+1}\left(\zeta\stackrel[n=1]{N}{\sum}\left(\rho_{n}^{k}e^{j\phi_{n}^{k}}\right)^{H}\right)\label{eq:78}
\end{equation}

\noindent where $\xi_{9}=\left|\mathbf{\bar{g}}_{b,r}\Theta\bar{\mathbf{g}}_{b,r}^{H}\right|^{2}$
and$\zeta=\mathbf{\bar{g}}_{b,r}\Theta\bar{\mathbf{g}}_{b,r}^{H}$.

\section*{Appendix D}

From $\mathcal{E}\left[R_{u_{2u}}\right]$ expression presented in
Theorem 4, we can find that 

\begin{equation}
p_{u_{2u}}=P_{b_{1}}R_{u_{th}}\frac{y_{2_{u_{2u}}}}{x_{1_{u_{2u}}}}+P_{b_{2}}R_{u_{th}}\frac{y_{2_{u_{2u}}}}{x_{1_{u_{2u}}}}+R_{u_{th}}\frac{V}{x_{1_{u_{2u}}}}+R_{u_{th}}\frac{\sigma_{b}^{2}}{x_{1_{u_{2u}}}}\label{eq:94}
\end{equation}

From $\mathcal{E}\left[R_{u_{2d}}\right]$ expression presented in
Theorem 2 we can find 

\begin{equation}
P_{b_{2}}=R_{d_{th}}P_{b_{1}}\frac{x_{1_{u_{2d}}}}{x_{1_{u_{2d}}}}+R_{d_{th}}p_{u_{1u}}\frac{y_{1_{u_{2d}}}}{x_{1_{u_{2d}}}}+R_{d_{th}}p_{u_{2u}}\frac{y_{2_{u_{2d}}}}{x_{1_{u_{2d}}}}+R_{d_{th}}\frac{\sigma_{u_{2d}}^{2}}{x_{1_{u_{2d}}}}
\end{equation}

\begin{equation}
P_{b_{2}}=P_{b_{1}}a_{1}+R_{d_{th}}p_{u_{1u}}a_{2}+a_{3}
\end{equation}

where $a_{1}=\frac{\left(\frac{R_{d_{th}}x_{1_{u_{2d}}}}{x_{1_{u_{2d}}}}+R_{u_{th}}R_{d_{th}}\frac{y_{2_{u_{2d}}}}{x_{1_{u_{2d}}}}\frac{y_{2_{u_{2u}}}}{x_{1_{u_{2u}}}}\right)}{\left(1-R_{u_{th}}R_{d_{th}}\frac{y_{2_{u_{2d}}}}{x_{1_{u_{2d}}}}\frac{y_{2_{u_{2u}}}}{x_{1_{u_{2u}}}}\right)}$,
$a_{2}=\frac{y_{1_{u_{2d}}}}{x_{1_{u_{2d}}}\left(1-R_{u_{th}}R_{d_{th}}\frac{y_{2_{u_{2d}}}}{x_{1_{u_{2d}}}}\frac{y_{2_{u_{2u}}}}{x_{1_{u_{2u}}}}\right)}$
and

$a_{3}=\frac{\eta}{\left(1-R_{u_{th}}R_{d_{th}}\frac{y_{2_{u_{2d}}}}{x_{1_{u_{2d}}}}\frac{y_{2_{u_{2u}}}}{x_{1_{u_{2u}}}}\right)}$.
Since $p_{u_{1u}}=P_{t}-\left(p_{u_{2u}}+\stackrel[i=1]{2}{\sum}P_{b_{i}}\right)$,
we can get

\begin{equation}
P_{b_{2}}=P_{b_{1}}a_{1}+a_{2}R_{d_{th}}P_{t}-a_{2}R_{d_{th}}p_{u_{2u}}-a_{2}R_{d_{th}}P_{b_{1}}-a_{2}R_{d_{th}}P_{b_{2}}+a_{3}
\end{equation}

\begin{equation}
P_{b_{2}}\left(1+a_{2}R_{d_{th}}\right)=P_{b_{1}}\left(a_{1}-a_{2}R_{d_{th}}\right)+a_{2}R_{d_{th}}P_{t}-a_{2}R_{d_{th}}p_{u_{2u}}+a_{3}\label{eq:93}
\end{equation}

Substituting (\ref{eq:94}) into (\ref{eq:93}) we get 

\begin{equation}
P_{b_{2}}=P_{b_{1}}b_{1}+P_{t}b_{2}-b_{3}\label{eq:107}
\end{equation}

\noindent where $b_{1}=\frac{\left(\left(a_{1}-a_{2}R_{d_{th}}\right)-a_{2}R_{d_{th}}R_{u_{th}}\frac{y_{2_{u_{2u}}}}{x_{1_{u_{2u}}}}\right)}{\left(\left(1+a_{2}R_{d_{th}}\right)+a_{2}R_{d_{th}}R_{u_{th}}\frac{y_{2_{u_{2u}}}}{x_{1_{u_{2u}}}}\right)},b_{2}=\frac{a_{2}R_{d_{th}}}{\left(\left(1+a_{2}R_{d_{th}}\right)+a_{2}R_{d_{th}}R_{u_{th}}\frac{y_{2_{u_{2u}}}}{x_{1_{u_{2u}}}}\right)}$
and

\noindent $b_{3}=\frac{a_{4}}{\left(\left(1+a_{2}R_{d_{th}}\right)+a_{2}R_{d_{th}}R_{u_{th}}\frac{y_{2_{u_{2u}}}}{x_{1_{u_{2u}}}}\right)}$.
Now, the rate at user 1 to detect user 2 message is 

\begin{equation}
\mathcal{E}\left[\bar{R}_{u_{1d}\rightarrow u_{2d}}\right]\approx\log_{2}\left(1+\frac{P_{b_{2}}x_{1_{u_{1d}}}}{P_{b_{1}}x_{1_{u_{1d}}}+p_{u_{1u}}y_{1_{u_{1d}}}+p_{u_{2u}}y_{2_{u_{1d}}}+\sigma_{u_{1d}}^{2}}\right)
\end{equation}

we can find

\begin{equation}
P_{b_{2}}=P_{b_{1}}c_{1}+P_{t}c_{2}-c_{3}\label{eq:112}
\end{equation}

where $c_{1}=\frac{\left(\left(\hat{a_{1}}-\hat{a_{2}}R_{d_{th}}\right)-\hat{a_{2}}R_{d_{th}}R_{u_{th}}\frac{y_{2_{u_{2u}}}}{x_{1_{u_{2u}}}}\right)}{\left(\left(1+\hat{a_{2}}R_{d_{th}}\right)+\hat{a_{2}}R_{d_{th}}R_{u_{th}}\frac{y_{2_{u_{2u}}}}{x_{1_{u_{2u}}}}\right)},c_{2}=\frac{\hat{a_{2}}R_{d_{th}}}{\left(\left(1+\hat{a_{2}}R_{d_{th}}\right)+\hat{a_{2}}R_{d_{th}}R_{u_{th}}\frac{y_{2_{u_{2u}}}}{x_{1_{u_{2u}}}}\right)}$,

$c_{3}=\frac{\hat{a_{4}}}{\left(\left(1+\hat{a_{2}}R_{d_{th}}\right)+\hat{a_{2}}R_{d_{th}}R_{u_{th}}\frac{y_{2_{u_{2u}}}}{x_{1_{u_{2u}}}}\right)}$,
$\hat{a_{1}}=\frac{\left(\frac{R_{d_{th}}x_{1_{u_{1d}}}}{x_{1_{u_{1d}}}}+R_{u_{th}}R_{d_{th}}\frac{y_{2_{u_{1d}}}}{x_{1_{u_{1d}}}}\frac{y_{2_{u_{2u}}}}{x_{1_{u_{2u}}}}\right)}{\left(1-R_{u_{th}}R_{d_{th}}\frac{y_{2_{u_{1d}}}}{x_{1_{u_{1d}}}}\frac{y_{2_{u_{2u}}}}{x_{1_{u_{2u}}}}\right)}$, 

$\hat{a_{2}}=\frac{y_{1_{u_{1d}}}}{x_{1_{u_{1d}}}\left(1-R_{u_{th}}R_{d_{th}}\frac{y_{2_{u_{1d}}}}{x_{1_{u_{1d}}}}\frac{y_{2_{u_{2u}}}}{x_{1_{u_{2u}}}}\right)}$,
$\hat{a_{3}}=\frac{\hat{\eta}}{\left(1-R_{u_{th}}R_{d_{th}}\frac{y_{2_{u_{1d}}}}{x_{1_{u_{1d}}}}\frac{y_{2_{u_{2u}}}}{x_{1_{u_{2u}}}}\right)}$

$\hat{a_{4}}=\hat{a_{2}}R_{d_{th}}R_{u_{th}}\frac{V}{x_{1_{u_{2u}}}}-\hat{a_{2}}R_{d_{th}}R_{u_{th}}\frac{\sigma_{b}^{2}}{x_{1_{u_{2u}}}}+\hat{a_{3}}$,
and

$\hat{\eta}=R_{u_{th}}R_{d_{th}}\frac{y_{2_{u_{1d}}}}{x_{1_{u_{1d}}}}\frac{V}{x_{1_{u_{2u}}}}+R_{u_{th}}R_{d_{th}}\frac{y_{2_{u_{1d}}}}{x_{1_{u_{1d}}}}\frac{\sigma_{b}^{2}}{x_{1_{u_{2u}}}}+R_{d_{th}}\frac{\sigma_{u_{2d}}^{2}}{x_{1_{u_{1d}}}}$.
Now, from (\ref{eq:107}) and (\ref{eq:112})

\begin{equation}
P_{b_{1}}b_{1}+P_{t}b_{2}-b_{3}=P_{b_{1}}c_{1}+P_{t}c_{2}-c_{3}
\end{equation}

and 

\begin{equation}
P_{b_{1}}=P_{t}\frac{\left(c_{2}-b_{2}\right)}{\left(b_{1}-c_{1}\right)}-\frac{\left(c_{3}-b_{3}\right)}{\left(b_{1}-c_{1}\right)}
\end{equation}

Substitute into (\ref{eq:107}) or (\ref{eq:112}) we can get

\begin{equation}
P_{b_{2}}=P_{t}\left(b_{1}\frac{\left(c_{2}-b_{2}\right)}{\left(b_{1}-c_{1}\right)}+b_{2}\right)-b_{1}\frac{\left(c_{3}-b_{3}\right)}{\left(b_{1}-c_{1}\right)}-b_{3}
\end{equation}

By substituting into (\ref{eq:94}) we can get

\begin{equation}
p_{u_{2u}}=P_{t}\left(\frac{\left(c_{2}-b_{2}\right)}{\left(b_{1}-c_{1}\right)}R_{u_{th}}\frac{y_{2_{u_{2u}}}}{x_{1_{u_{2u}}}}+\left(b_{1}\frac{\left(c_{2}-b_{2}\right)}{\left(b_{1}-c_{1}\right)}+b_{2}\right)R_{u_{th}}\frac{y_{2_{u_{2u}}}}{x_{1_{u_{2u}}}}\right)+c_{4}
\end{equation}
where
$c_{4}=-\frac{\left(c_{3}-b_{3}\right)}{\left(b_{1}-c_{1}\right)}R_{u_{th}}\frac{y_{2_{u_{2u}}}}{x_{1_{u_{2u}}}}
-b_{1}\frac{\left(c_{3}-b_{3}\right)}{\left(b_{1}-c_{1}\right)}R_{u_{th}}\frac{y_{2_{u_{2u}}}}{x_{1_{u_{2u}}}}-b_{3}R_{u_{th}}\frac{y_{2_{u_{2u}}}}{x_{1_{u_{2u}}}}+R_{u_{th}}\frac{V}{x_{1_{u_{2u}}}}+R_{u_{th}}\frac{\sigma_{b}^{2}}{x_{1_{u_{2u}}}}$.

The power expressions in (96)-(98) are the same as (37)-(39), after re-naming the variables.

\bibliographystyle{IEEEtran}
\bibliography{bib}

\begin{thebibliography}{10}
\providecommand{\url}[1]{#1}
\csname url@samestyle\endcsname
\providecommand{\newblock}{\relax}
\providecommand{\bibinfo}[2]{#2}
\providecommand{\BIBentrySTDinterwordspacing}{\spaceskip=0pt\relax}
\providecommand{\BIBentryALTinterwordstretchfactor}{4}
\providecommand{\BIBentryALTinterwordspacing}{\spaceskip=\fontdimen2\font plus
\BIBentryALTinterwordstretchfactor\fontdimen3\font minus \fontdimen4\font\relax}
\providecommand{\BIBforeignlanguage}[2]{{%
\expandafter\ifx\csname l@#1\endcsname\relax
\typeout{** WARNING: IEEEtran.bst: No hyphenation pattern has been}%
\typeout{** loaded for the language `#1'. Using the pattern for}%
\typeout{** the default language instead.}%
\else
\language=\csname l@#1\endcsname
\fi
#2}}
\providecommand{\BIBdecl}{\relax}
\BIBdecl

\bibitem{ref1}
M.~Di~Renzo, A.~Zappone, M.~Debbah, M.-S. Alouini, C.~Yuen, J.~de~Rosny, and S.~Tretyakov, ``Smart radio environments empowered by reconfigurable intelligent surfaces: How it works, state of research, and the road ahead,'' \emph{IEEE Journal on Selected Areas in Communications}, vol.~38, no.~11, pp. 2450--2525, 2020.

\bibitem{Reef1}
C.~Pan, H.~Ren, K.~Wang, J.~F. Kolb, M.~Elkashlan, M.~Chen, M.~Di~Renzo, Y.~Hao, J.~Wang, A.~L. Swindlehurst, X.~You, and L.~Hanzo, ``Reconfigurable intelligent surfaces for 6g systems: Principles, applications, and research directions,'' \emph{IEEE Communications Magazine}, vol.~59, no.~6, pp. 14--20, 2021.

\bibitem{https://doi.org/10.48550/arxiv.2301.00276}
\BIBentryALTinterwordspacing
A.~Salem, K.-K. Wong, and C.-B. Chae, ``Impact of phase-shift error on the secrecy performance of uplink {RIS} communication systems,'' 2023. [Online]. Available: \url{https://arxiv.org/abs/2301.00276}
\BIBentrySTDinterwordspacing

\bibitem{Dai20}
L.~Dai \emph{et~al.}, ``Reconfigurable intelligent surface-based wireless communications: Antenna design, prototyping, and experimental results,'' \emph{IEEE Access}, vol.~8, no.~1, pp. 45\,913--45\,923, 2020.

\bibitem{Ref22}
S.~Zhang and R.~Zhang, ``Capacity characterization for intelligent reflecting surface aided {MIMO} communication,'' \emph{IEEE Journal on Selected Areas in Communications}, vol.~38, no.~8, pp. 1823--1838, 2020.

\bibitem{Ref33}
J.~Zhang, J.~Liu, S.~Ma, C.-K. Wen, and S.~Jin, ``Large system achievable rate analysis of ris-assisted {MIMO} wireless communication with statistical {CSIT},'' \emph{IEEE Transactions on Wireless Communications}, vol.~20, no.~9, pp. 5572--5585, 2021.

\bibitem{Ref44}
K.~Xu, J.~Zhang, X.~Yang, S.~Ma, and G.~Yang, ``On the sum-rate of ris-assisted {MIMO} multiple-access channels over spatially correlated {Rician} fading,'' \emph{IEEE Transactions on Communications}, vol.~69, no.~12, pp. 8228--8241, 2021.

\bibitem{Ref55}
K.~Zhi, C.~Pan, H.~Ren, and K.~Wang, ``Power scaling law analysis and phase shift optimization of ris-aided massive {MIMO} systems with statistical csi,'' \emph{IEEE Transactions on Communications}, vol.~70, no.~5, pp. 3558--3574, 2022.

\bibitem{refone}
X.~Mu, Y.~Liu, L.~Guo, J.~Lin, and R.~Schober, ``Simultaneously transmitting and reflecting (star) ris aided wireless communications,'' \emph{IEEE Transactions on Wireless Communications}, vol.~21, no.~5, pp. 3083--3098, 2022.

\bibitem{mag1}
Y.~Liu, X.~Mu, J.~Xu, R.~Schober, Y.~Hao, H.~V. Poor, and L.~Hanzo, ``Star: Simultaneous transmission and reflection for 360° coverage by intelligent surfaces,'' \emph{IEEE Wireless Communications}, vol.~28, no.~6, pp. 102--109, 2021.

\bibitem{Ref2}
B.~Zhao, C.~Zhang, W.~Yi, and Y.~Liu, ``Ergodic rate analysis of {STAR-RIS} aided {NOMA} systems,'' \emph{IEEE Communications Letters}, vol.~26, no.~10, pp. 2297--2301, 2022.

\bibitem{Ref3}
H.~Liu, G.~Li, X.~Li, Y.~Liu, G.~Huang, and Z.~Ding, ``Effective capacity analysis of {STAR-RIS}-assisted {NOMA} networks,'' \emph{IEEE Wireless Communications Letters}, vol.~11, no.~9, pp. 1930--1934, 2022.

\bibitem{Ref4}
A.~Papazafeiropoulos, Z.~Abdullah, P.~Kourtessis, S.~Kisseleff, and I.~Krikidis, ``Coverage probability of {STAR-RIS}-assisted massive {MIMO} systems with correlation and phase errors,'' \emph{IEEE Wireless Communications Letters}, vol.~11, no.~8, pp. 1738--1742, 2022.

\bibitem{Ref5}
Z.~Xie, W.~Yi, X.~Wu, Y.~Liu, and A.~Nallanathan, ``{STAR-RIS} aided {NOMA} in multicell networks: A general analytical framework with gamma distributed channel modeling,'' \emph{IEEE Transactions on Communications}, vol.~70, no.~8, pp. 5629--5644, 2022.

\bibitem{Ref6}
H.~Ma, H.~Wang, H.~Li, and Y.~Feng, ``Transmit power minimization for {STAR-RIS}-empowered uplink {NOMA} system,'' \emph{IEEE Wireless Communications Letters}, vol.~11, no.~11, pp. 2430--2434, 2022.

\bibitem{Ref7}
F.~Fang, B.~Wu, S.~Fu, Z.~Ding, and X.~Wang, ``Energy-efficient design of {STAR-RIS} aided {MIMO-NOMA} networks,'' \emph{IEEE Transactions on Communications}, vol.~71, no.~1, pp. 498--511, 2023.

\bibitem{Ref8}
J.~Chen and X.~Yu, ``Ergodic rate analysis and phase design of {STAR-RIS} aided {NOMA} with statistical {CSI},'' \emph{IEEE Communications Letters}, vol.~26, no.~12, pp. 2889--2893, 2022.

\bibitem{Ref9}
J.~Zuo, Y.~Liu, Z.~Ding, L.~Song, and H.~V. Poor, ``Joint design for simultaneously transmitting and reflecting {(STAR) RIS} assisted {NOMA} systems,'' \emph{IEEE Transactions on Wireless Communications}, vol.~22, no.~1, pp. 611--626, 2023.

\bibitem{ref9a}
J.~Xu, Y.~Liu, X.~Mu, R.~Schober, and H.~V. Poor, ``Star-riss: A correlated t and r phase-shift model and practical phase-shift configuration strategies,'' \emph{IEEE Journal of Selected Topics in Signal Processing}, vol.~16, no.~5, pp. 1097--1111, 2022.

\bibitem{ref9b}
X.~Yue, J.~Xie, Y.~Liu, Z.~Han, R.~Liu, and Z.~Ding, ``Simultaneously transmitting and reflecting reconfigurable intelligent surface assisted noma networks,'' \emph{IEEE Transactions on Wireless Communications}, vol.~22, no.~1, pp. 189--204, 2023.

\bibitem{Chae1}
M.~Chung, M.~S. Sim, J.~Kim, D.-K. Kim, and C.-B. Chae, ``Prototyping real-time full duplex radios,'' \emph{IEEE Communications Magazine}, vol.~53, no.~9, pp. 56--64, Sep 2020.

\bibitem{Chae2}
M.~S. Sim, M.~Chung, D.~Kim, J.~Chung, and C.-B. Kim, D. K. andand~Chae, ``Nonlinear self-interference cancellation for full duplex radios: From link- and system-level performance perspectives,'' \emph{IEEE Communications Magazine}, vol.~55, no.~9, pp. 158--167, Sep. 2017.

\bibitem{Chae3}
M.~Chung, D.~K. Sim, M. S.~andKim, and C.-B. Chae, ``Compact full duplex {MIMO} radios in {D2D} underlaid cellular networks: From system design to prototype results,'' \emph{IEEE Communications Magazine}, vol.~5, pp. 16\,601--16\,617, Sep. 2017.

\bibitem{Ref12}
Y.~Cai, M.-M. Zhao, K.~Xu, and R.~Zhang, ``Intelligent reflecting surface aided full-duplex communication: Passive beamforming and deployment design,'' \emph{IEEE Transactions on Wireless Communications}, vol.~21, no.~1, pp. 383--397, 2022.

\bibitem{Ref13}
D.~Xu, X.~Yu, Y.~Sun, D.~W.~K. Ng, and R.~Schober, ``Resource allocation for irs-assisted full-duplex cognitive radio systems,'' \emph{IEEE Transactions on Communications}, vol.~68, no.~12, pp. 7376--7394, 2020.

\bibitem{ref14}
Z.~Peng, Z.~Zhang, C.~Pan, L.~Li, and A.~L. Swindlehurst, ``Multiuser full-duplex two-way communications via intelligent reflecting surface,'' \emph{IEEE Transactions on Signal Processing}, vol.~69, pp. 837--851, 2021.

\bibitem{ref15}
P.~Guan, Y.~Wang, H.~Yu, and Y.~Zhao, ``Joint beamforming optimization for ris-aided full-duplex communication,'' \emph{IEEE Wireless Communications Letters}, vol.~11, no.~8, pp. 1629--1633, 2022.

\bibitem{Ref17}
P.~K. Sharma, N.~Sharma, S.~Dhok, and A.~Singh, ``Ris-assisted fd short packet communication with non-linear eh,'' \emph{IEEE Communications Letters}, vol.~27, no.~2, pp. 522--526, 2023.

\bibitem{b1}
B.~Smida, A.~Sabharwal, G.~Fodor, G.~C. Alexandropoulos, H.~A. Suraweera, and C.-B. Chae, ``Full-duplex wireless for {6G}: Progress brings new opportunities with new challenges,'' \emph{IEEE Journal on Selected Areas in Communications}, vol.~41, no.~9, pp. 2729--2750, Sep. 2023.

\bibitem{b2}
------, ``Guest editorial: Full duplex and its applications,'' \emph{IEEE Journal on Selected Areas in Communications}, vol.~41, no.~9, pp. 2725--2728, Sep. 2023.

\bibitem{SI1}
E.~Everett, A.~Sahai, and A.~Sabharwal, ``Passive self-interference suppression for full-duplex infrastructure nodes,'' \emph{IEEE Transactions on Wireless Communications}, vol.~13, no.~2, pp. 680--694, 2014.

\bibitem{SI2}
E.~Ahmed and A.~M. Eltawil, ``All-digital self-interference cancellation technique for full-duplex systems,'' \emph{IEEE Transactions on Wireless Communications}, vol.~14, no.~7, pp. 3519--3532, 2015.

\bibitem{bidirectional2}
T.~J. Oechtering and M.~Skoglund, ``Bidirectional broadcast channel with random states noncausally known at the encoder,'' \emph{IEEE Transactions on Information Theory}, vol.~59, no.~1, pp. 64--75, 2013.

\bibitem{mrc}
S.~Khisa, M.~Almekhlafi, M.~Elhattab, and C.~Assi, ``Full duplex cooperative rate splitting multiple access for a miso broadcast channel with two users,'' \emph{IEEE Communications Letters}, vol.~26, no.~8, pp. 1913--1917, 2022.

\bibitem{powerallocation1}
Y.~Sun, D.~W.~K. Ng, Z.~Ding, and R.~Schober, ``Optimal joint power and subcarrier allocation for full-duplex multicarrier non-orthogonal multiple access systems,'' \emph{IEEE Transactions on Communications}, vol.~65, no.~3, pp. 1077--1091, 2017.

\bibitem{powerallocation2}
A.~Abrardo, M.~Moretti, and F.~Saggese, ``Power and subcarrier allocation in 5g noma-fd systems,'' \emph{IEEE Transactions on Wireless Communications}, vol.~19, no.~12, pp. 8246--8260, 2020.

\bibitem{RefMain}
A.~Papazafeiropoulos, P.~Kourtessis, and I.~Krikidis, ``{STAR-RIS} assisted full-duplex systems: Impact of correlation and maximization,'' \emph{IEEE Communications Letters}, vol.~26, no.~12, pp. 3004--3008, 2022.

\bibitem{salim11}
M.~Alouini and A.~J. Goldsmith, ``Area spectral efficiency of cellular mobile radio systems,'' \emph{IEEE Transactions on Vehicular Technology}, vol.~48, no.~4, pp. 1047--1066, July 1999.

\bibitem{booknew}
A.~M. Mathai, \emph{An Introduction to Geometrical Probability}, Gordon and Breach Science Publishers.

\bibitem{book2}
M.~Abramowitz and I.~A. Stegun, \emph{Handbook of Mathematical Functions With Formulas, Graphs, and Mathematical Tabl}, Washington,D.C.: U.S. Dept. Commerce, 1972.

\end{thebibliography}

\end{document}